\documentclass[epj]{svjour}
\usepackage[pdftex]{graphicx}
\usepackage{dcolumn}
\usepackage{bm}
\usepackage{color}
\usepackage{longtable}
\usepackage{xr}
\usepackage{subfigure}

\newcommand{\beq}{\begin{equation}}
\newcommand{\eeq}{\end{equation}}
\newcommand{\beqa}{\begin{eqnarray}}
\newcommand{\eeqa}{\end{eqnarray}}

\newcommand{\bem}{\begin{math}}
\newcommand{\eem}{\end{math}}
\newcommand{\rar}{{\rightarrow}}
\newcommand{\Rar}{{\Rightarrow}}
\newcommand{\bfr}{{\bf r}}
\newcommand{\bfu}{{\bf u}}
\newcommand{\bfp}{{\bf p}}

\newcommand{\bff}{{\bf f}}
\newcommand{\bfv}{{\bf v}}
\newcommand{\bfe}{{\bf e}}

\newcommand{\bfF}{{\bf F}}

\newcommand{\bsigma}{{\boldsymbol \sigma}}

\newcommand{\bomega}{{\boldsymbol \omega}}

\def\strutdepth{\dp\strutbox}
\def\nw#1{\strut\vadjust{\kern-\strutdepth\vtop to0pt{\vss\hbox to\hsize
{\hskip\hsize\hskip5pt$\leftarrow$\hss\strut}}}{\em #1}}
\usepackage{multirow}
\usepackage{fixltx2e}

\def\tll#1{\textcolor{black}{#1}} 

\def\tlll#1{\textcolor{black}{#1}} 

\newcommand{\pdiff}[2]{\frac{\partial #1}{\partial #2}}

\newcommand{\cosT}{\cos \theta}
\newcommand{\sinT}{\sin \theta}

\def\mean#1{\left< #1 \right>}

\def\NY#1{\textcolor{black}{#1}}
\def\NYY#1{\textcolor{black}{#1}}
\def\NYYY#1{\textcolor{black}{#1}}

\usepackage{amsmath,amssymb}	

\begin{document}
\title{From hydrodynamic lubrication to many-body interactions in dense suspensions of active swimmers}
\author{Natsuhiko Yoshinaga\inst{1,2,3} \and Tanniemola B. Liverpool\inst{3,4,5}
}                     
\offprints{}          
\institute{
WPI - Advanced Institute for Materials Research, Tohoku University,
Sendai 980-8577, Japan, 
\email{yoshinaga@tohoku.ac.jp}
\and
MathAM-OIL, AIST,
Sendai 980-8577, Japan
\and
The Kavli Institute for Theoretical Physics, University of California,
Santa Barbara, CA 93106, USA
\and
School of Mathematics, University of Bristol, Bristol, BS8 1TW, UK, 
\email{t.liverpool@bristol.ac.uk}
\and
BrisSynBio, Life Sciences Building, Tyndall Avenue, Bristol, BS8 1TQ, UK
}
\date{Received: date / Revised version: date}
%
\abstract{
We study how hydrodynamic interactions affect 
  the collective behaviour of  active particles suspended in a fluid at high concentrations, with particular attention to lubrication forces which appear when the particles are very close to one another. We compute exactly the limiting behaviour of the hydrodynamic interactions between two spherical (circular) active swimmers in very close
  proximity to one another in the general setting in both three and (two) dimensions.  Combining this with far-field interactions, we develop a novel numerical scheme which allows us to study the collective behaviour of {\em large} numbers of active particles {\em with}
  accurate hydrodynamic interactions when close to one another.  We study active swimmers whose intrinsic flow fields are characterised by force dipoles and quadrupoles. Using this scheme, we are able to show that lubrication forces when the particles are very close to each other can play as important a role as long-range hydrodynamic interactions in determining their many-body behaviour.   \tlll{We find that when the swimmer force dipole is large, finite clusters and open
  gel-like clusters appear rather than complete phase
  separation.} 
  This suppression is  due to near-field lubrication interactions.
For swimmers with small force dipoles, we find surprisingly that a globally polar ordered phase
  appears because near field lubrication rather than long-range hydrodnamics dominate the alignment mechanism.
  Polar order is present for very large system sizes
  and is stable to fluctuations with a finite noise amplitude. We explain the 
  emergence of polar order using a minimal model in which only the leading
  rotational effect of the near-field interaction is included. These phenomena are also reproduced in two dimensions. 
%
} 
\titlerunning{Hydrodynamic lubrication in dense suspension of  active swimmers}
\authorrunning{N. Yoshinaga and T. B. Liverpool}
\maketitle

\section{Introduction}

Active materials are condensed matter systems which contain  components that are 
self-driven out of equilibrium
, and have been
 studied as inspiration for 
new smart materials and as a framework to understand 
aspects of  cell motility~\cite{Marchetti:2013,Toner2005,ramaswamy:2010}.  
They are characterised by a plethora of fascinating non-equilibrium collective phenomena such as swirling, alignment,
 pattern formation, dynamic cluster formation and phase separation~\cite{vicsek:1995,Henricus:2012,Cates:2012,Palacci:2013,Bricard:2013}  which have recently generated much interest.
This phenomenology has inspired models of  active suspensions involving assemblies of actively moving objects in far-from-equilibrium states at various length and time scales ranging from
animals to cells and microorganisms.
Theoretical descriptions of active systems range from continuum
 models~\cite{Marchetti:2013,Bertin:2013a} to discrete collections of
 self-propelled active particles~\cite{vicsek:1995}.
A recent influential classification of self-propelled active particle systems has  grouped them into 
 {\it dry} and {\it wet} systems~\cite{Marchetti:2013}.
 Dry systems do not have momentum conserving dynamics.
 Examples include Vicsek models~\cite{vicsek:1995,Bertin:2013a} and
 Active Brownian particle (ABP) models interacting via soft {\em
 repulsive} potentials~{\cite{Fily:2012,Buttinoni:2013,Redner:2013,Speck:2016}}
 Wet systems conserve momentum via a coupling to a fluid leading to hydrodynamic interactions between active particles.
Microorganism suspensions are a typical example of this class.
Inspired by these systems, several models have been proposed such as squirmers driven by surface deformations~\cite{Lighthill:1952,blake:1971} and Janus particles driven by surface chemical reactions~\cite{golestanian:2005}, which under certain conditions can be mapped to squirmers with tangential deformations~\cite{Ibrahim:2016}.
While dry models are attractive due to their computational tractability,
it is not clear under which conditions they are useful models for the
increasing number of  experiments on micron sized self-propelled
particles~\cite{Palacci:2013,Bricard:2013}.
Hence it is essential to clarify the role of hydrodynamic interactions.
In this work, we systematically construct equations of
motion for wet active particles, namely, the dynamics of their position and
orientation.

\tll{
When hydrodynamic interactions are present, the motion of a particle is affected by long-range
interactions from other particles due to both fluid flow and pressure.
This results in many-body effects due to higher-order multipoles
\NYYY{(see sect.~\ref{sec.comparison.ref})}.
In addition, even
the two-body interaction between particles in close proximity
(near-field) has non-trivial singular behaviour, requiring either exquisitely fine meshes
between the gap or prior knowledge of analytical solutions in this region ~\cite{ishikawa:2006,Llopis:2006,Ishikawa:2008a,Ishimoto:2013,Li:2014,SharifiMood:2015,Papavassiliou:2016}.
It remains a significant challenge to overcome these technical hurdles and hence obtain
an understanding of collective
behaviour ~\cite{Mucha:2004}.
Because numerical simulations with hydrodynamics require significantly
more computational power (for example, $\mathcal{O}(N^3)$ for Stokesian
Dynamics, and $\mathcal{O}((L/b)^d)$ for Lattice Boltzmann method, where
$N$, $L$, $b$, and \NYYY{$d$} are the number of particles, the system size, the
size of a mesh, and \NYYY{space dimension}, respectively), studies of these systems to date contain 
relatively small number of particles compared with studies of ABPs and
the Vicsek model 
\cite{Molina:2013,Zoettl:2014,Matas-Navarro:2014,Delfau:2016}.
 It has been noted however that in those dry active systems,  macroscopic behaviour, such
 as the spontaneous breaking of symmetries and the emergence of global order, are strongly affected by finite size effects.
 This leads one to worry if studies based on small numbers of particles can be accurately used to study macroscopic behaviour of wet active systems.
As a result, the far-field approximation  
~\cite{Spagnolie:2012} has often been used to account for
hydrodynamic interactions \cite{Saintillan:2015,Stenhammar:2017}.
Within this approximation, swimmers are represented as point dipoles
(denoted as $v_2$ in this work) and quadrupoles (denoted by $v_1$).
This approximation while able to study large numbers of active particles only makes sense  for describing very dilute suspensions because the far-field approximation is no longer valid 
 when the swimmers are close to one another. 
 So for non-dilute suspensions, there is a need for studies of wet active systems which take account of hydrodynamics accurately for small swimmer separations and can also deal with large number of particles.
}

\begin{figure*}[htbp]
 \begin{center}
  \resizebox{0.99\textwidth}{!}{%
  \includegraphics{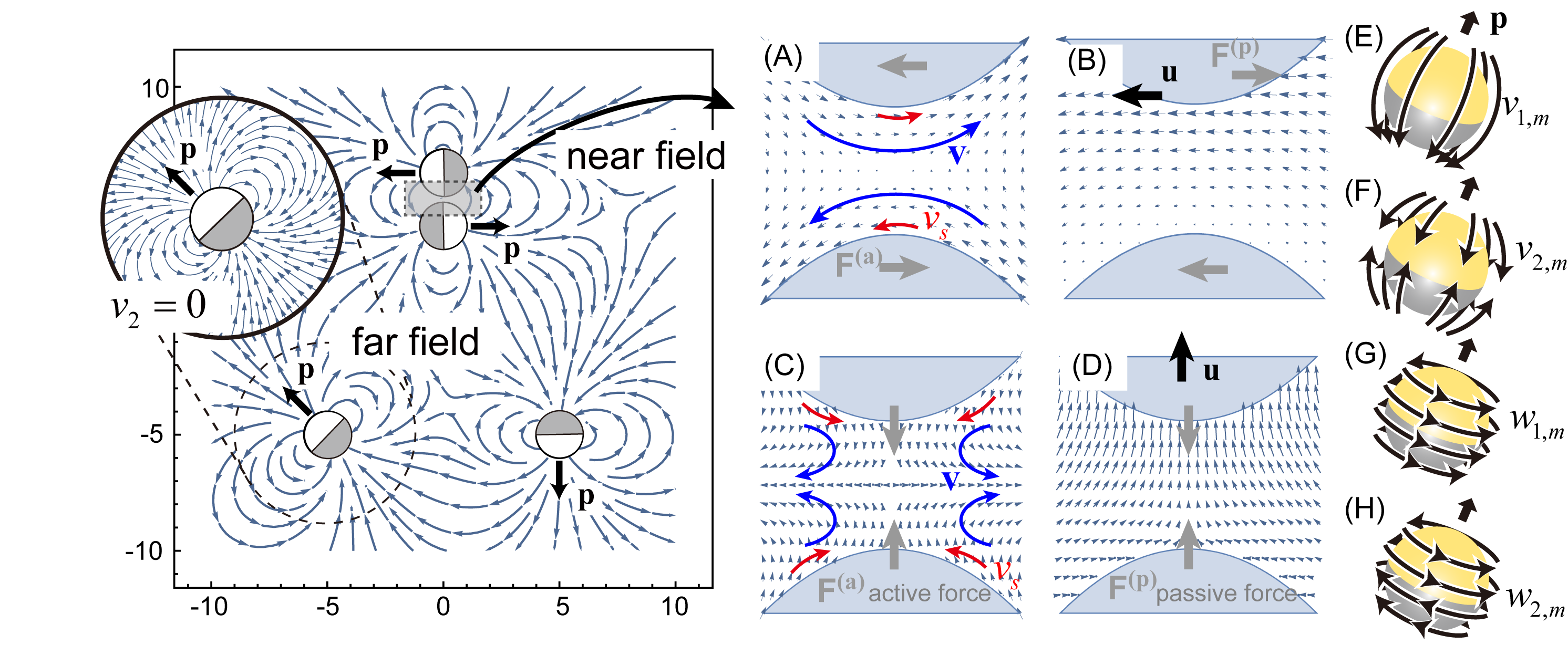}
}
 \caption{
 \tll{
 (Colour Online) (Left) Schematic interaction between swimmers and
 surrounding flow.
 Each particle creates a leading order multipolar flow, which perturbs
 the translational and rotational motion of others in quite different ways in the far-field (when swimmers are well separated) and near-field (in close proximity to one another).
 (Right)  The near-field flow field around shearing (A,B) and converging (C,D) swimmers.
The active problems are shown in (A) and (C), and the passive problems
 are shown in (B) and (D).
 The direction of motion ${\bf u}$ and the surface slip flow are indicated by
 thick black and thin red arrows, respectively.
 The force ${\bf F}^{(a)}$ and ${\bf F}^{(p)}$ acting on each swimmer in those configurations is shown by
 a gray arrow for active and passive problems, respectively.
 As a guide to the eye, the blue arrows show fluid flow.
  \tlll{
(E-H) Schematic of the slip velocity on the surface of a swimmer and the flow it generates. The arrows show the tangential flow in the
  direction,  ${\bf \Psi}_{lm} (\theta,\varphi)$ for $l=1$ (E) and
  $l=2$ (F), and in the direction, $ {\bf \Phi}_{lm} (\theta,\varphi)$
  for $l=1$ (G) and $l=2$ (H)
  in \NY{eq}.(\ref{janus.intro.slip}).
  }
 \label{fig.schematics}
 }
}
 \end{center}
\end{figure*}

We have recently proposed a model including both long-range interactions and
accurate near-field lubrication forces at the level of two-body interactions,
sacrificing however, accuracy at intermediate length scales.
Using it we have studied and
demonstrated a number of new classes of collective behaviours of squirmers in large systems
\cite{Yoshinaga:2017a}.
\tll{
An outline of our method is schematically illustrated in fig.~\ref{fig.schematics}. We study active swimmers (squirmers) which when isolated generate an intrinsic flow in the far-field that are equivalent to those characterised by force dipoles and quadrupoles, the magnitudes of which, can be controlled as microscopic parameters.
When two particles are far apart, the flow field is dominated by
the far-field interaction, which is a flow field generated by an
isolated swimmer with perturbation due to the other swimmer.
When two particles approach each other, their translational and rotational motion
is dominated by the near-field interaction.
The near-field motion and rotation are obtained by decomposing the flow
field into its passive (see figs.~\ref{fig.schematics}(B) and (D)) and
active (see figs.~\ref{fig.schematics}(A) and (C)) parts and solving for the exact hydrodynamic flow fields in the lubrication limit.
In this work, we will discuss the details of our new methodology and present more precise analyses of the results produced by the model.
In addition, we present a detailed comparison between our method and 
other methods to handle hydrodynamic interactions in active materials.
We point out a number of aspects of the problem that are sometimes overlooked.
We also present a detailed analysis of the mechanisms of aggregation of the active swimmers. This involves introducing and characterising quantitatively, dynamic and static clusters, which appear
when the absolute value of the dipole force is large $|v_2| \gg 0$.
 When $|v_2| \simeq 0$,  a globally polar ordered phase appears as reported in 
 \cite{Yoshinaga:2017a} and 
 \cite{ishikawa:2006,Alarcon:2016}.
 Here, we confirm this result by showing polar order is robust
 to fluctuations of finite amplitude. Further more we are able to identify its origin and show that a global polar phase appears in a 
 simplified model where only the rotational near-field interaction and
 steric interactions are included. 
Finally, our model is applied to two-dimensional squirmers, that is,
particles move in the two-dimensional plane with purely two-dimensional hydrodynamic 
interactions.
We also present  general and explicit formulas for the far-field and near-field fluid flows, which,
to our knowledge, have not been reported  elsewhere before.
}

This paper is organised as follows:
In sect.~\ref{sec.model}, we describe our model squirmers.
The flow field generated by an isolated squirmer is obtained exactly.
Then, the interactions between two squirmers are analytically
calculated in two asymptotic limits; when the two swimmers  (i) are far apart and (ii)
are nearly touching.
A comparison to previous studies is also presented.
Readers who are interested only in the results of analytical
calculations of the interactions may skip the beginning of the section
and go directly to \NY{eqs.}(\ref{trans.vel}), (\ref{ang.vel}), and the following discussions. 
We use the result of the interactions in sect.~\ref{sec.model}B to perform 
 numerical simulations of the collective behaviour of many squirmers  in
 sect.~\ref{sec.collective}.
We also consider two-dimensional systems, which are discussed in
sect.~\ref{sec.2D}.
In sect.~\ref{sec.collision}, the dynamics of a collision of a pair of squirmers are analysed
in detail from their initial approach to their eventual separation.
We conclude with sect.~\ref{sec.summary}, which summarizes our
results.
The technical details are outlined in the Appendices.


\section{Model}
\label{sec.model}

Each particle (squirmer) is characterized by its position and orientation $({\bf
r}^{(i)} , {\bf p}^{(i)} )$ with dynamics given by
 \begin{align}
\dot{{\bf r}}^{(i)} 
&=
{\bf u}^{(i)} 
\label{janus.dotr.u}
\\
\dot{{\bf p}}^{(i)} 
&=
{\bm \omega}^{(i)} 
\times {\bf p}^{(i)}
\label{janus.dotp.omega} 
\end{align}
The translational and angular velocities of each particle are denoted by ${\bf u}^{(i)} $ and
${\bm \omega}^{(i)} $, respectively.
%
To obtain the velocities, we solve for the fluid mediated interaction between all the squirmers.
The fluid is taken as incompressible in the vanishing Re limit :
\begin{align}
 \eta \nabla ^2 {\bf v} - \nabla p 
 &=
 0 \, 
 \label{eq:stokes1} \\
 \; \;
 \nabla \cdot {\bf v}
 &=
 0
\label{eq:stokes2}\end{align}
where $\eta$ is viscosity, ${\bf v}({\bf r})$ is the
velocity, and $p ({\bf r})$ the pressure. The boundary
condition on the swimmer surface is a sum of rigid translational, ${\bf u}$ and rotational, ${\bm \omega}$
motion and an active slip flow, ${\bf v}_s$ driving self-propulsion:
\begin{align}
 \left. {\bf v} \right|_{{\bf r} = {\bf R}} &=
{\bf u} + {\bm \omega} \times {\bf R} + {\bf v}_s \label{BC} \\
  {\bf v}_s &= \sum_{l \ge 1,m} \left[ v_{lm} {\bf \Psi}_{lm} (\theta,\varphi) + w_{lm} {\bf \Phi}_{lm} (\theta,\varphi)
\right] \quad ,
\label{janus.intro.slip}
\end{align}
for a swimmer with centre at the origin.
The fluid velocity vanishes
 at infinity, ${\bf
v}|_{r \rightarrow \infty} = 0$,
with $\theta$ the angle with the $z$-axis and $\varphi$ with the $x$-axis on the $xy$-plane.
The slip velocity ${\bf v}_s$ can be very efficiently expanded in 
vector spherical harmonics, ${\bf \Psi}_{lm}$ and ${\bf \Phi}_{lm}$
which span the space of tangent vectors on the sphere \cite{Hill:1954} (as done in \NY{eq.~}(\ref{janus.intro.slip})). 
The vector spherical harmonics  ${\bf Y}_{lm} (\theta,\varphi)$ , ${\bf \Phi}_{lm} (\theta,\varphi)$ and ${\bf
\Psi}_{lm}(\theta,\varphi)$ are defined by
\begin{align}
 {\bf Y}_{lm} (\theta,\varphi)
&=
\hat{{\bf r}} Y_l^m(\theta,\varphi)
\label{vector.spherical.Y}
\\
{\bf \Psi}_{lm}(\theta,\varphi)
&=
r \nabla Y_l^m(\theta,\varphi)
\label{vector.spherical.Psi}
\\
{\bf \Phi}_{lm}(\theta,\varphi)
&=
{\bf r} \times \nabla Y_l^m (\theta,\varphi)
\label{vector.spherical.Phi}
,
\end{align}
where $Y_l^m(\theta,\phi)$  
are the Laplace spherical harmonics.
The spherical harmonics, $Y_l^m(\theta,\varphi)$ are defined here as
\begin{align}
 Y_l^m(\theta,\varphi)
&=
\mathcal{N}_{lm}
P_l^m (\cos \theta) e^{i m \varphi}
\end{align}
with $P_l^m (\cos \theta)$, the associated Legendre polynomial  of degree $l$
and order $m$.
We use the normalization convention
$\mathcal{N}_{lm}=\sqrt{\frac{(2l+1)(l-m)!}{4\pi(l+m)!}}$.

\begin{figure}[htbp]
 \begin{center}
    \resizebox{0.5\textwidth}{!}{%
  \includegraphics{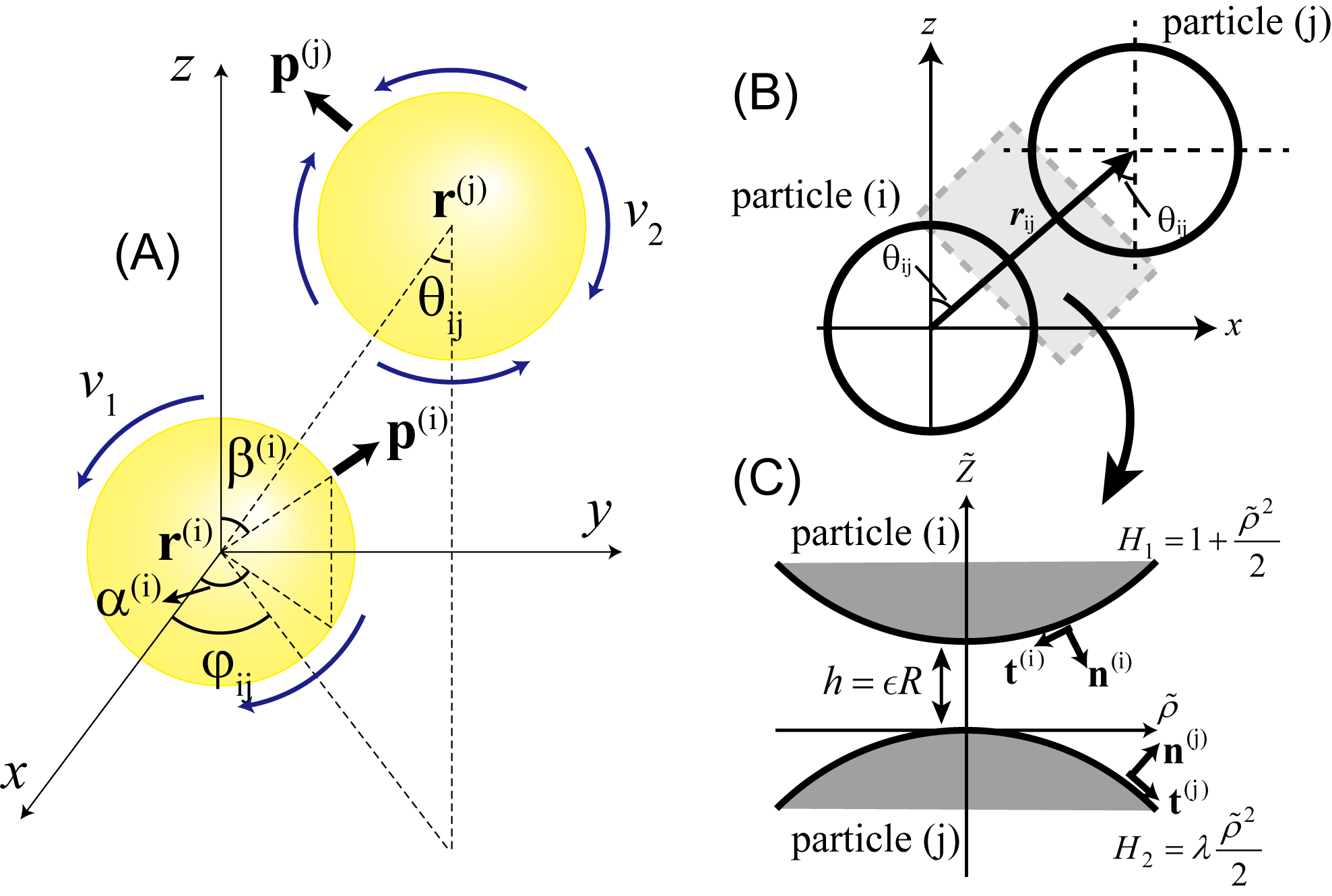}
}
 \caption{
 \NYY{
  (A) Characterising the state of the swimmer with position/orientation
    $\bf r, p$ and its relation to the parameters  $v_1,v_2,\alpha,\beta$ and the polar angles $\theta,\varphi$.
 (B) The coordinate for arbitrary position of the $i$th and $j$th
 particles. (C) The stretched coordinate used for calculation of the
 near-field interaction between the spheres (i) and (j) placed along
 the $z$-axis.
 }
\label{fig.sm.coordinate}
}
 \end{center}
\end{figure}

\tlll{
The second term in \NY{eq.~}(\ref{janus.intro.slip}) characterised by its
coefficient $w_{lm}$ represents rotational slip
associated with spinning motion (see \NY{fig.}~\ref{fig.schematics}(G-H)).
 This does not, however, mean $w_{lm}=0$ ensures axissymetric flow
 around the particle even when it is isolated.
We discuss this issue below \NY{eq.~}(\ref{uniaxial.slip.corrsp}).
For simplicity, in this work, we neglect this term in the following 
and from now on set $w_{lm}=0$.
}
The  swimmer axis
\begin{align}
 {\bf p}
 &=
 \left(
\cos \alpha \sin \beta,
\sin \alpha \sin \beta,
\cos \beta
 \right)
 \label{swimmer.axis}
\end{align}
 is a unit vector with azimuthal, $\alpha$
 and polar, $\beta$ angles \NYY{(see fig.~\ref{fig.sm.coordinate}(A)
 and its projection onto the $xz$ plane shown in fig.~\ref{fig.sm.coordinate}(B))}.
 %

For uniaxial particles, $v_{lm}$ can, without loss of generality, be decomposed into a magnitude $v_l$
and the angles ($\alpha$, $\beta$) as $v_{lm} = \mathcal{D}_{m0}(\alpha,\beta) v_l
$ with Wigner matrix $\mathcal{D}_{mm'}(\alpha,\beta)$
rotating a $2l+1$-dimensional vector of the
$v_{lm}$~\cite{edmonds:1957}.
This is explicitly expressed as
\begin{align}
 v_{lm}
 &=
 \sqrt{\frac{4 \pi}{2l+1}} Y_l^{m*} (\beta,\alpha) v_l \; , 
 \label{uniaxial.slip.corrsp}
\end{align}
in terms of the modes $v_l$.
\NY{
Here, $Y_l^{m*} (\beta,\alpha)$ is the complex conjugate of $Y_l^{m} (\beta,\alpha)$.
}
Note that in the double sum in \NY{eq.~}(\ref{janus.intro.slip}), for each $l$; $m$ takes values $-l,-l+1,\ldots, 0, \ldots, l-1, l$ and hence $v_{lm}$ can be represented by a $2l+1$-dimensional vector.
For $l=1$ and $l=2$, the vectors are shown in
Appendix~\ref{sec.appendix.SH}.
A uniaxial squirmer is characterised by the strength of
the $l$-th mode, $v_l$, and its swimming orientation by the two angles
($\alpha$ and $\beta$).
When the particle is not uniaxial, the $v_{lm}$ for $l \geq 2$ will have additional
degrees of freedom.

The flow field and resulting motion for an isolated squirmer is
summarized in the next section.
The first, $l=1$ mode in \NY{eq.~}(\ref{janus.intro.slip}) is a quadrupolar 
flow  with strength $v_1$ (i.e. it decays like $r^{-3}$ at large distances $r$ from the swimmer, detailed analysis shows that it is a potential flow source/sink dipole)~\cite{blake:1974} leading to translational motion while the second, $l=2$ mode is
a dipole  of strength $v_2$ (decays like $r^{-2}$)~\cite{blake:1974}.
\tlll{
In much of the literature on squirmers~\cite{Lighthill:1952,blake:1971},
the surface slip flow ${\bf v}_s$ in \NY{eq.~}(\ref{janus.intro.slip}) is
expanded in terms of Legendre polynomials assuming an 
axisymmetric flow profile, with coefficients $B_l$ which can be obtained using the orthogonality of the Legendre polynomials.
We can relate the parameters $B_l$ of these papers, to the coefficients $v_l$ of our expansion in terms of the vector spherical harmonics:}
\begin{align}
 B_l
 &=
 - v_l \sqrt{\frac{(2l+1)}{4\pi}}
 \frac{l(l+1)}{2}
\end{align}
leading to
$v_1=-\sqrt{4\pi/3}B_1$ and $v_2 = (-2/3)\sqrt{\pi/5} B_2 $, and
thus the squirmer parameter $\beta_{2/1} $($B_2/B_1 =\beta$ in the conventional
notation) is $\beta_{2/1} = (3 \sqrt{5/3}) v_2/v_1 \simeq 3.87 v_2/v_1  $.

\subsection{Flow field around an isolated squirmer}

In this section, we give the explicit forms of the flow field generated
by an isolated {\em non-spinning} squirmer in terms of the vector spherical harmonics.
The velocity field and the pressure field are in general expressed as
\begin{align}
{\bf v}({\bf r}) 
&=
\sum_{l \ge 1,m}
\left[
f_{lm}^{(t)} (r)
{\bf \Psi}_{lm} (\theta,\varphi)
+
f_{lm}^{(n)} (r)
{\bf Y}_{lm} (\theta,\varphi)
\right]
\label{janus.vel.vector.spherical}
\\
p({\bf r})
&=
\sum_{l \ge 1,m}
P_{lm}
\left(
\frac{R}{r}
\right)^{l+1}
 Y_l^m (\theta,\varphi)
 .
\end{align}
The coefficient functions $f_{lm}^{(t)}$, $f_{lm}^{(n)}$, and $P_{lm}$ are
obtained by solving the Stokes equation (\ref{eq:stokes1},\ref{eq:stokes2}) with the effective slip boundary
conditions \NY{eq.}(\ref{janus.intro.slip}).

The coefficients for the first mode, $l=1$  are
\begin{align}
f_{1,m}^{(n)}
 &=
u_m \left(
\frac{R}{r}
\right)^3
\\
f_{1,m}^{(t)}
 &=
- \frac{1}{2}
u_m \left(
\frac{R}{r}
\right)^3
\\
P_{1,m}
 &= 0
 .
\end{align}
For the higher modes, $l \geq 2$, they are
\begin{align}
f_{lm}^{(n)} 
&=
\frac{l(l+1)}{2} v_{lm}
\left[
\left(
\frac{R}{r}
\right)^l
- 
\left(
\frac{R}{r}
\right)^{l+2}
\right]
\\
f_{lm}^{(t)} 
&=
-\frac{v_{lm}}{2}
\left[
(l-2)
\left(
\frac{R}{r}
\right)^l
- 
l
\left(
\frac{R}{r}
\right)^{l+2}
\right]
\\
P_{lm}
&=
\frac{\eta}{R}
l(2l-1) v_{lm}
\\
u_m
&=
- \frac{2}{3} v_{1,m}
.
\end{align}
The self-propulsion velocities of the particle are expressed in Cartesian coordinates as
\begin{align}
{\bf u} 
&=
\sum_m 
u_m
\left[
{\bf Y}_{1,m} (\theta,\varphi)
+ {\bf \Psi}_{1,m} (\theta,\varphi)
\right]
\nonumber \\ &=
\begin{pmatrix}
\mathcal{N}_{1,1} (-u_1 + u_{-1}) \\
i \mathcal{N}_{1,1}  (- u_1 - u_{-1}) \\
\mathcal{N}_{1,0} u_{0}
\end{pmatrix}
 .
\label{isolated.u}
 \\
 {\bm \omega} &= 0 \; .
 \label{isolated.omega}
\end{align}
It is noteworthy that (1) the isolated squirmer is force and torque free
by construction due to the absence of the $1/r$ tern in the $l=1$ modes and (2) the self-propulsion speed depends only on the first, $l=1$ mode and is
 independent of the higher modes, $l \geq 2$.
Since we do not consider self-rotating squirmers, the angular velocity of the isolated swimmer is identically zero.



We use the vector spherical harmonics for expansion of an active slip
flow and a surrounding flow.
This is because two tangential flows associating with translation and
self-rotation become clearer (See \NY{eq.~}(\ref{janus.intro.slip})). 
 Nevertheless, it is possible to express these flow in terms of
 Cartesian tensors.
 In Appendix~\ref{sec.Cartesian}, we demonstrate an explicit translation of
 a fluid flow around an isolated swimmer into the form of Cartesian tensors.

\subsection{Pairwise interactions between squirmers}
\label{sec.pariwise}

When two squirmers are present, the flow field generated by one will affect the other and hence lead to modification of the self-propulsion velocities. To calculate the modified flow and hence the effective pairwise hydrodynamic interactions between two squirmers, we solve the Stokes equation (\ref{eq:stokes1},\ref{eq:stokes2}) with slip boundary conditions, \NY{eqs.~}(\ref{janus.intro.slip}) on the surface of both swimmers taking into account both of their orientations (denoted by the unit vector ${\bf p}^{(i)}$) which are chosen arbitrarily.
This gives rise to modified velocities, ${\bf u}^{(i)}, {\bm \omega}^{(i)}$ for each squirmer which will also depend on the position and orientation of the other.
To compute the velocities, ${\bf u}^{(i)}, {\bm \omega}^{(i)}$  
we split the problem into two parts, a force and torque acting
on the sphere with: 1st,  slip boundary conditions without
translational and rotational motion, and 2nd  with the non-slip boundary
conditions undergoing rigid-body motion ${\bf u}^{(i)}$ and ${\bm \omega}^{(i)}$.
We shall call the former the active force (torque), ${\bf F}^{(a)} ({\bf T}^{(a)})$ and the latter the
passive force (torque), ${\bf F}^{(p)} ({\bf T}^{(p)})$.
The force and torque-free conditions imply, 
\begin{align}
 {\bf F}^{(a)} + {\bf F}^{(p)} &=0
 \label{force.balance}
\\
 {\bf T}^{(a)} + {\bf T}^{(p)} &=0.
  \label{torque.balance}
\end{align}
The passive force and torque are proportional to ${\bf u}^{(i)}$ and ${\bm
\omega}^{(i)}$ while the active force and torque are proportional to $v_{lm}$.
Therefore the force and torque-free conditions are exactly expressed by
 \begin{align}   {\bf L}^{(i,j)}   \cdot  \begin{pmatrix} {\bf u}^{\NYY{(j)}} \\
{\bm \omega}^{\NYY{(j)}} \\
\end{pmatrix}
  &=  \mathcal{L}^{(i,j)}_{lm}   \cdot v_{lm}^{(j)}
\label{squirmer.v.omega.to.vlm}
 \end{align}
where the superscript $(i)$ denotes $i$th particle.
${\bf L}$ is called the passive resistance matrix, likewise
$\mathcal{L}$, the active resistance matrix,
The resistance matrices depend only on the shapes of the particles and
their relative positions (${\bf r}_{ij}$ being the vector between their centers, see fig.~\ref{fig.sm.coordinate}(A)).
The matrices ${\bf L},\mathcal{L}$ can be calculated exactly 
for pairs of particles in two asymptotic limits : (1)
when their separation, $h_{ij}=r_{ij}-2R, r_{ij}=|{\bf r}_{ij}|$ is much
less than their radius \NYY{(near-field, fig.~\ref{fig.sm.coordinate}(C))} and  (2) when their separation
is much greater than their radius \NYY{(far-field, fig.~\ref{fig.sm.coordinate}(B))}.
There is long history of calculation of the passive matrix~
\cite{Jeffrey:1984,Kim:1991}. 
Here we  compute 
the active resistance matrix for both far-field and
near-field  in the general setting.
Previous near field results 
have been obtained only for axisymmetric surface flow-fields. We note that non-axisymmetric flow fields around {\em isolated} swimmers, however have been considered before, e.g. in \cite{Pak:2014}.

\begin{figure*}[htbp]
 \begin{center}
    \resizebox{0.99\textwidth}{!}{%
  \includegraphics{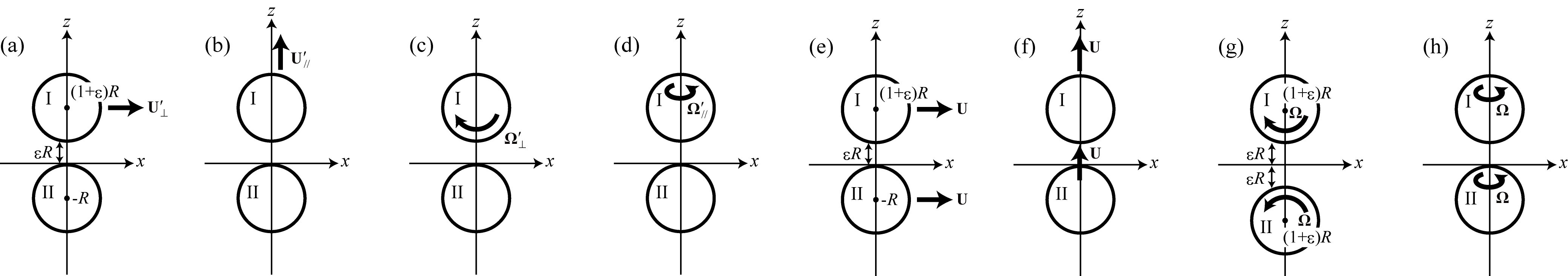}
}
\caption{
The eight configurations for translational and rotational motion of two
 particles.
The thick arrows show the direction of motion.
\label{fig.sec1.8problems}
}
 \end{center}
\end{figure*}

It should be noted that to obtain the velocity and angular
velocity for swimmers oriented in arbitrary directions, one must compute \NY{eqs.~}(\ref{force.balance}) and (\ref{torque.balance}) for  {\it all} possible orientation 
directions. Hence, below, we decompose the problem into the eight possible motions (see 
fig.~\ref{fig.sec1.8problems})  and calculate the velocities for two arbitrarily oriented swimmers  in \NY{eqs.~}(\ref{trans.vel}) and (\ref{ang.vel}).
This has not been achieved before; we will compare our
analytical results with previous studies in sect.~\ref{sec.comparison.ref}.

The equations
(\ref{janus.dotr.u}), (\ref{janus.dotp.omega}), and
(\ref{squirmer.v.omega.to.vlm})  form a closed complete dynamical system.
%
We use a general form for the velocities valid in both 
far and near field limits:
  \NYY{
 \begin{align}
{\bf u}^{(i)}
 &=
 {\bf u}_0^{(i)} \lambda^{(i)} +
 \sum_{j \neq i, l,m}
 \left[
 u_{lm,\parallel}^{(j)}
 {\bf Y}_{lm}^{(ji)} 
 +
u_{lm,\perp}^{(j)}
 {\bf \Psi}_{lm}^{(ji)} 
 \right]
 \label{trans.vel}
 \\
{\bm \omega}^{(i)} 
&=
 \sum_{j \neq i, l,m}
\omega_{lm}^{(j)}
 {\bm \Phi}_{lm}^{(ji)}
  .
 \label{ang.vel}
 \end{align}
 }
The isolated squirmer moves with the velocity
\begin{align}
 {\bf u}_0^{(i)} &= u_0 {\bf p}^{(i)}
 \nonumber \\
 u_0 &=
 - \frac{2}{3} \sqrt{\frac{3}{4\pi}} v_1
 \label{squirmer.u0}
 .
\end{align}
$\lambda^{(i)}=1$ when the $i$th particle is away from near-field region
of any other particles and $\lambda^{(i)}=0$ otherwise.
For a pair of squirmers (labelled $i,j$) with arbitrary positions (and orientation), we can define a set of spherical coordinates with relative positions ${\bf r}_{ij}$ and relative angles $\theta_{ij}$ and
$\varphi_{ij}$ \NYY{(see fig.~\ref{fig.sm.coordinate})}.
 We denote $ {\bf Y}_{lm} ( \theta_{ji}, \varphi_{ji}) = {\bf
 Y}_{lm}^{(ji)}$ and $
\theta_{ji} 
=
\pi - \theta_{ij}
$ and $
\varphi_{ji}
=
\pi + \varphi_{ij}
$.
$\theta_{ij}$ is the polar angle of the vector between the centers of
the $i$th and $j$th particles and $\varphi_{ij}$ is its azimuthal
angle.

The symmetry of this interaction is obtained from the following property of
the spherical harmonics: 
\begin{align}
Y_{l}^m (\theta_{ji},\varphi_{ji}) 
&=
(-1)^{l} Y_{l}^m (\theta_{ij},\varphi_{ij}) 
.
\end{align}
Similarly, the vector spherical harmonics transform as
\begin{align}
{\bf Y}_{lm} (\theta_{ji},\varphi_{ji}) 
=
 (-1)^{l+1} {\bf Y}_{lm} (\theta_{ij},\varphi_{ij})
 \\
{\bf \Psi}_{lm} (\theta_{ji},\varphi_{ji}) 
=
(-1)^{l+1}
 {\bf \Psi}_{lm} (\theta_{ij},\varphi_{ij}) 
\\
{\bf \Phi}_{lm} (\theta_{ji},\varphi_{ji}) 
=
(-1)^{l}
 {\bf \Phi}_{lm} (\theta_{ij},\varphi_{ij}) .
\end{align}

\subsubsection{Near-field}

The calculation of the near-field interaction is a conceptually straightforward, if technically complicated computation. 
For completeness, we briefly outline the main steps in this section.
Without loss of generality, two spherical squirmers are
placed along the $z$-axis (fig.~\ref{fig.sec1.8problems}).
Then flow induced by the interaction is decomposed into parallel and perpendicular
directions corresponding to the ${\bf Y}_{lm}$ and ${\bm \Psi}_{lm}$ terms.
Since we may arbitrarily choose the $x$ and $y$ axes, we have two translational velocities $u_x$ and $u_z$ and two angular
velocities $\omega_y$ and $\omega_z$ for each particle.
These eight unknown velocities are determined by the eight configurations shown in fig.~\ref{fig.sec1.8problems}.
In the suite, we neglect (d) and (h) in fig.~\ref{fig.sec1.8problems} since we do not consider
spinning motion.

When the two squirmers almost touch each other, we may assume the gap
between two sphere surfaces $h$ is much smaller than their radius (see fig.~\ref{fig.sm.coordinate}(C)).
We introduce the small parameter
\begin{align}
\epsilon 
&=
\frac{h}{R}
.
\end{align}

The asymptotic behaviour  in the limit of small $\epsilon$ for the dynamics of rigid spheres moving slowly past each other (which we call the passive problem) has been studied by a number of authors~\cite{Majumdar:1967,ONeill:1969,ONeill:1970,Cooley:1971,Cooley:1969,Jeffrey:1982,Jeffrey:1984a}.   Here we adapt and extend these results to the problem of squirmers in close proximity (the active problem). The leading order contribution in $\epsilon$ to the force and torque depends on the particular problem (configuration) considered (see fig.~\ref{fig.sec1.8problems}).
It is well known that for the passive problem with the configuration
(a)-(c) in fig.~\ref{fig.sec1.8problems}, the force and
torque diverges with decreasing separation between the spheres, $\epsilon$; $F^{(p)} \sim 1/\epsilon$ for (b) \cite{Jeffrey:1982} and $F^{(p)} \sim
\log \epsilon$ for (a) and (c) \cite{ONeill:1970,Jeffrey:1984a}.
For (e) and (f), both force and torque do not diverge as $\epsilon
\rightarrow 0$ and thus $F^{(p)} \sim \mathcal{O}(\epsilon^0)$ and
$T^{(p)} \sim \mathcal{O}(\epsilon^0)$ \cite{ONeill:1969,Cooley:1969}.
For (g), the force is regular $F^{(p)} \sim \mathcal{O}(\epsilon^0)$
while the torque is singular $T^{(p)} \sim
\mathcal{O}(\log \epsilon)$ \cite{ONeill:1969}.
Although we have not considered spinning motion in this article - problems (d) and (h) in
fig.~\ref{fig.sec1.8problems}, both the problems are not singular and
thus $T^{(p)} \sim \mathcal{O}(\epsilon^0)$ for (d)
\cite{Jeffrey:1984a,Cooley:1971} and (h) \cite{Majumdar:1967}.

When $\epsilon \rightarrow 0$, it is convenient to use the following stretched coordinate $(X,Y,Z)$ (see fig.~\ref{fig.sm.coordinate}(C))
\begin{align}
x  
&=
\sqrt{\epsilon} X
\\
y
&=
\sqrt{\epsilon} Y
\\
z &= \epsilon Z
.
\end{align}
We scale lengths with $R$ defining, dimensionless coordinates $\tilde{X}=X/R$,
$\tilde{Y}=Y/R$, and $\tilde{Z}=Z/R$.
It is convenient to use cylindrical coordinates $(\rho,\varphi,z)$, 
where $\rho = x^2 + y^2$,  or equivalently 
stretched cylindrical coordinates $(\tilde{\rho}, 
\varphi, \tilde{Z})$ with $\tilde{\rho} = \sqrt{\tilde{X}^2 + \tilde{Y}^2}$ . 
We can without loss of generality place the centres of the two  spheres (identified by labels 1,2)  at  (see fig.~\ref{fig.sm.coordinate}(B))
\begin{align}
{\bf r}_{G}^{(1)} 
&=
(0,0,(1+\epsilon) R)
\\
{\bf r}_{G}^{(2)} 
&=
(0,0,-R)
.
\end{align}
The surfaces of the spheres are given by the boundary surfaces $H_1$ and $H_2$ respectively.
The position at the boundary $\tilde{Z} = H_1$ and $\tilde{Z}=H_2$ is
\begin{align}
H_1
&=
1 + \frac{1}{2} \tilde{\rho}^2
+ \mathcal{O}(\epsilon)
=
H_0
+ \mathcal{O}(\epsilon)
\\
H_2
&=
- \frac{1}{2} \tilde{\rho}^2
+ \mathcal{O}(\epsilon)
.
\end{align}

The squirmer  slip velocity boundary condition,  $\bf{v}_s (\theta,\varphi)$ given in \NY{eq.~}(\ref{janus.intro.slip}) can be expanded 
 in the coordinate system above.
In the near field, the boundary condition is expanded as a series in $\epsilon$.
The details of the calculation of the passive and active forces of
the configuration (b) in fig.~\ref{fig.sec1.8problems} are outlined in
Appendix~\ref{sec.app.axisymmetric.3D} and (a), (c), (g) in Appendix~\ref{sec.app.nonaxisymmetric.3D}.

The passive force and torque are linear functions of translational
and rotational velocity, while the active force and torque are expressed
in terms of the surface slip velocity.
Using the generalized vector
$
 {\bf U}
 =
 \left(
 {\bf u}^{(1)}, {\bf u}^{(2)},
 {\bm \omega}^{(1)}, {\bm \omega}^{(2)}
 \right)
$
and
$
 {\bf V}_{s,lm}
 =
 \left(
 v_{lm}^{(1)}, 
 v_{lm}^{(2)} 
 \right)
$, 
the problem is rewritten as
\begin{align}
 {\bf L} \cdot {\bf U}
 &=
 -
 \sum_{l,m}   \mathcal{L}_{lm} 
 \cdot {\bf V}_{s,m}
\end{align}
where ${\bf L}$ is an invertible $4d \times 4d$ matrix while $\mathcal{L}_{lm}$ is $4d
\times 2(2l+1)$ matrix.
The passive and
active resistance
matrices ${\bf L}$ and $\left. \mathcal{L}_{lm} \right.$ are obtained from the analyses  of  the previous
sections. 
The passive resistance
matrix must be  inverted to compute the velocities and angular velocities of the swimmers.
Following \cite{Swan:2011}, we include the diagonal terms corresponding to the force and torque of an
translating and rotating isolated particle.

In the near field limit, our lubrication analysis enables us to obtain all the singular contributions as $\epsilon$ approaches zero.
They arise from relative motion between the particles that give either shear
or converging flow as shown in figs.~\ref{fig.schematics}(A)-(D).
A similar analysis can be performed for motion perpendicular to the
line between two centers.
This corresponds to the problems (a), (c), (e), and (g).


When the separation
between two sphere surfaces, $h_{ij}=\epsilon R, \epsilon \ll 1$, is small compared to their size (see fig.~\ref{fig.sm.coordinate} (B) and fig.~\ref{fig.schematics}),
we systematically extend the passive lubrication calculation for nearby spheres with non-slip boundary conditions~\cite{ONeill:1967,COOLEY:1968}
 to account for active slip boundary conditions of squirmers
and hence obtain coefficients in \NY{eqs.~}(\ref{trans.vel}) and (\ref{ang.vel}) as
\begin{align}
 u^N_{lm,\parallel}
 &=
 - \epsilon \log \epsilon V_{lm}
 + \frac{1}{3} W_{1,m} \delta_{l,1}
  \label{near.para}
 \\
 u^N_{lm,\perp}
 &=
  \frac{1}{2}  V_{lm}
 + \frac{1}{3} W_{1,m} \delta_{l,1}
 \\
 \omega^N_{lm}
 &=
 -  \frac{2}{5 R} 
 W_{lm}
 \label{near.omega}
\end{align}
with
\begin{align}
 V_{lm} = \frac{l(l+1)}{2}
\left(
(-1)^l v_{lm}^{(i)} + v_{lm}^{(j)}
 \right)
\\
 W_{lm} =  \frac{l(l+1)}{2}
 \left(
 (-1)^l v_{lm}^{(i)} -  v_{lm}^{(j)}
 \right)
 .
\end{align}
The leading order velocity parallel to the center line is
$\mathcal{O}(\epsilon \log \epsilon)$.
This is due to the balance of the passive lubrication force $F^{(p)} \sim
u/\epsilon$ and the active force $F^{(a)} \sim v_{lm} \log \epsilon$, which
is singular but only logarithmic since the contribution
from incompressibility is small here.
This result is consistent with \cite{ishikawa:2006,Wuerger:2016}.
Perpendicular motion and rotation are 
$\mathcal{O}(1)$ because both passive and active forces are logarithmic.
Combining all of these, we obtain the result in \NY{eqs.~}(\ref{near.para})-(\ref{near.omega})
above.

\subsubsection{Far field}

The far-field interaction between the $i$th and $j$th particles follows from  
Faxen's laws ~ \cite{Kim:1991} giving 
coefficients in \NY{eqs.~}(\ref{trans.vel}) and (\ref{ang.vel}) as
\begin{align}
  u^F_{1,m,\parallel}
  &=
  -\frac{2}{3}
  \frac{1}{\tilde{r}_{ij}^{3}}
 v_{1,m}
 \label{far.para1}
 ,
 \\
  u^F_{2,m,\parallel}
  &=
   \frac{3}{\tilde{r}_{ij}^{2}}
  v_{2,m}
 ,
 \\
  u^F_{1,m,\perp}
  &=
 \frac{1}{3}
 \frac{1}{\tilde{r}_{ij}^{3}}
    v_{1,m}
  , \\
  \omega^F_{2,m}
  &=
   - \frac{3}{\tilde{r}_{ij}^{3}}
 v_{2,m}/R \;
\label{far.omega2}
 ,
\end{align}
 and coefficients $l \geq 3$ vanish.
Here $\tilde{r}_{ij} = r_{ij}/R$.
Notice that in this limit the first,
 $l=1$ mode cannot induce rotation since it generates a potential flow.
Hence for  spheres,  the first mode contributes to rotation only via
multi-scattering effects. 


\subsubsection{Implementation}

\tlll{To describe the active resistance matrix for arbitrary separations
between particles, we interpolate between two limiting expressions, the far-field
and near field,  using a $\tanh$ function centred at $r=2.5R$ with width $0.1R$, that is 
much smaller than the size of a particle}.
\tlll{
Note that \NY{eq.~}(\ref{near.para}) gives rise to an effective repulsive
interaction between two particles because the speed of convergence goes to zero as
the separation between the particle surfaces become small.
This is also true for passive particles, but the  speed goes 
to zero slower for active particles due to the factor of $\log
\epsilon$ in \NY{eq.~}(\ref{near.para}).
}
However due to fact that finite time steps are required for the numerical implementation of any multi-particle simulation, we require a short range repulsive interaction between all the particles to avoid overlap between them (i.e. to stop the separation between the centres becoming less than double the radius). 
In Stokesian Dynamics simulations, a short-range repulsive force, ${\bf F}_{ij}$ between swimmers $i$ and $j$, whose centres are separated by the vector ${\bf r}_{ij}$,  is added to the 
 equation of motion \NY{eq.~}(\ref{janus.dotr.u})
\cite{brady:1988} (see also \cite{Melrose:1995}), where
\begin{align}
{\bf F}_{ij} 
&=
F_0 \frac{ e^{-\epsilon/r_s}}{r_s (1- e^{-\epsilon/r_s}) }
 \hat{\bf r}_{ij}
 \label{potential.Brady}
\end{align}
where $\epsilon = r_{ij}/R - 2$ and 
$r_s = 1/227$ \cite{brady:1988}.
A similar repulsive interaction has also been included in simulations of squirmers
using the Boundary Element Method~\cite{ishikawa:2006,Spagnolie:2012}.
In this work, we experimented with a variety of repulsive interactions
(including the one described above) and found that while a particular
choice changes the positions of boundaries of different classes of
behaviour, they do not lead to qualitative changes of the behaviour observed. Hence to obtain the phase diagrams which are the main result of this paper we used the common and well-known truncated Lennard-Jones repulsive interaction, i.e. we we add a short-range repulsive force ${\bf F}_{ij} = -\partial U/\partial {\bf r_{ij}}$ to \NY{eq.~}(\ref{janus.dotr.u}) where
\begin{align}
U (r) 
 &=
 \begin{cases}
-2
\left(
\frac{2R}{r_{ij}}
 \right)^6
 +
 \left(
\frac{2R}{r_{ij}}
  \right)^{12}
  & \mbox{         for       }
  r_{ij} < 2R
  \\
  0
  &  \mbox{         otherwise      }
 \end{cases}
\label{repulsive.potential}
\end{align}
\tlll{
With this method, the particles overlap at most 2\% of their radius for
neutral swimmers at the maximum density that we used.
}

\subsection{Two-dimensional squirmers}
\label{sec.2D.squimers}

\NYY{
In two dimensions, the expansion of the slip velocity in terms of spherical harmonics is
replaced by one in terms of sines and cosines 
\begin{align}
v_s
 &=
 \sum_{m=1}^{\infty}
 \left[
 v_{s,m} \sin m\theta
 + \tilde{v}_{s,m} \cos m \theta
 \right]
 .
\end{align}
The radial and tangential direction are expressed in polar coordinates as
${\bf r}
 =
 \left(
\cosT, \sinT
 \right)
 $ and $
{\bf t}
 =
 \left(
-\sinT, \cosT
 \right)
$ (see fig.~\ref{fig.sm.coordinate}(B)).
Then, the flow field of an isolated squirmer is expressed in terms of
$
 {\bf r}_m
 =
 \left(
\cos m \theta, \sin m \theta
 \right)
 $ and 
 $
 {\bf t}_m
 =
 \left(
-\sin m \theta, \cos m \theta
 \right)
 $ (see also Appendix~\ref{sec.app.2D} for the explicit form of the flow
 field).
The velocity of the particle is
\begin{align}
{\bf u}_0
 &=
 - \frac{1}{2 \pi R}
 \int v_s {\bf t} dS
 =
 \frac{1}{2} (v_{s,1} {\bf e}_x - \tilde{v}_{s,1} {\bf e}_y)
 .
 \label{2D.isolated.vel}
\end{align}
}

\tll{
In three dimensions, the dynamics is governed by \NY{eqs.~}(\ref{trans.vel}) and
(\ref{ang.vel}) supplemented with \NY{eqs.~}(\ref{far.para1})-(\ref{far.omega2})
for far-field and \NY{eqs.~}(\ref{near.para})-(\ref{near.omega})
for near-field.
In two dimensions, the dynamics is governed by the following equations:
\begin{align}
{\bf u}^{(i)}
 &=
 {\bf u}_0^{(i)} \lambda^{(i)} +
 \sum_{j \neq i, m}
 \left[
 u_{m,\parallel}^{(j)}
\hat{\bf r}_{ji}
 +
u_{m,\perp}^{(j)}
\hat{\bf t}_{ji}
 \right]
 \label{trans.vel.2D}
 \\
\omega^{(i)} 
&=
 \sum_{j \neq i, m}
\omega_{m}^{(j)}
 ,
 \label{ang.vel.2D}
\end{align}
where  expressions for $u_{m,\parallel}, u_{m,\perp}$ in the near-field and far-field regimes are given in Appendix~\ref{sec.app.2D} and 
\begin{align}
\hat{\bf r}_{ij,m}
 &=
 \left(
\cos m \theta_{ij}, \sin m \theta_{ij}
 \right)
\\
 \hat{\bf t}_{ij,m}
 &=
 \left(
 -\sin m\theta_{ij}, \cos m\theta_{ij}
 \right).
\end{align}
\NYY{
In order to compute the interaction, we consider six problems in (a)-(c) and (e)-(g) in fig.~\ref{fig.sec1.8problems}.
}
The explicit expressions are given in Appendix~\ref{sec.app.2D}.
In contrast with three-dimensional systems, the leading order velocity
parallel to the center line is $\mathcal{O}(\sqrt{\epsilon})$.
Perpendicular motion and rotation are the same as three-dimensional
squirmers and proportional to $\mathcal{O}(1)$.
}

\subsection{Comparison to other work}
\label{sec.comparison.ref}

In this section, we compare the methods developed in this paper and previous
work.
First, we summarise the existing approaches.
There are two methods commonly used  to compute near-field interactions between spherical objects: 
bispherical (bipolar in two dimensions) coordinates and matched
asymptotic analyses.
Pair-wise interactions computed using bispherical coordinates are exact
but the solution is expressed in terms of an infinite algebraic system of equations.
 The sum may be evaluated numerically with truncation after a finite number of terms, with more and more
 terms necessary as the two spheres approach each other~\cite{YARIV:2003}.
 For large separations between spheres, this truncation  is justified and
 the far-field interaction is reproduced.
 
For the near-field interaction, there is no general method to express this
 infinite series as an expansion in terms of the small separation $\epsilon$.
 We are only aware of one case in which this has been done, for perfectly aligned 
 converging spheres 
 (problem (b) in fig.~\ref{fig.sec1.8problems}) in which the
 series for solid spheres is approximated for small $\epsilon$ ~\cite{Cox:1967} (also used in \cite{Papavassiliou:2016} for a squirmer
 problem).
 The matched asymptotic analysis (used in this work) is an expansion of the
 solution for small $\epsilon$. Its results are, of course, not exact for arbitrary separation but become asymptotically 
 exact when $\epsilon \ll 1$.
 In this particular case, the two approaches are identical both in the scaling with 
 $\epsilon$ and the prefactor of the leading asymptotic term for $\epsilon \ll 1$
 \cite{Cox:1967,Cooley:1969a}.

The matched asymptotic method is appealing because it removes the need for cumbersome calculations of very many terms of the series obtained using 
 the bispherical coordinate method. The analysis here, is based on the expansion of inner and outer regions.
 The inner region is obtained using the stretched coordinate system shown in
 fig.~\ref{fig.sm.coordinate} and the outer region is obtained using the
 tangent-sphere coordinate system~\cite{Cox:1967,Cooley:1969a}.
 The solution of each coordinate is matched at an appropriate limit.
 Clearly obtaining the full expansion will be as cumbersome as using 
 bispherical coordinates. The point is that the leading order terms in the asymptotic expansion are a very good approximation to the full solution 
 when the spheres are very close to each other. Calculating the leading order terms is relatively straightforward~\cite{ONeill:1967} and hence this gives a very 
 efficient and accurate calculation of the interactions for spheres in
 close proximity.
 In this paper, we treat only the leading-order singularities which diverge as $\epsilon\;  \rar \;  0$, while
 for the finite $\mathcal{O}(1)$ terms, we use the values valid for isolated swimmers.
 This is because it is known that the $\mathcal{O}(1)$ terms
 obtained by matched asymptotics are not particularly accurate and
 typically these terms have been obtained either using numerical estimates of the
 exact expression using bispherical coordinates or from direct
 numerical solutions of the Stokes equation \cite{Jeffrey:1984} which give values comparable to those of isolated swimmers.

In \cite{Papavassiliou:2016}, an {exact} solution of the Stokes equation
was computed for a axisymmetric arrangement of two squirmers using bispherical coordinates
and the reciprocal theorem. 
However this work is of limited use for the study of collective behaviour (the subject of this paper) because it by construction addresses only a highly reduced set of interaction scenarios, those with axisymmetric flow fields. To study the collective behaviour one must be able to describe {\em all} possible directions of approach and orientations of the two swimmers.
Since the system was assumed to be axisymmetric, only the motion and rotation
corresponding to (b), (d), (f), and (h) in fig.~\ref{fig.sec1.8problems}
{\em can be} considered.
In order to obtain the equation of motion of each swimmer, we need the 
functional form of the interaction for  {\it all} possible directions.
In the current work, we do not consider a spinning motion corresponding
to $w_{lm} \neq 0$ in \NY{eq.~}(\ref{janus.intro.slip}), and therefore do not study rotational
motion - problems  (d) and (h) in fig.~\ref{fig.sec1.8problems}.
Only the translational motion studied in problems (b) and (f) in fig.~\ref{fig.sec1.8problems} can be addressed within the 
framework of \cite{Papavassiliou:2016}. However, as there is no explicit formula discussed in this paper, a
direct comparison is not possible.

Bispherical coordinates were also used in \cite{SharifiMood:2015} to
compute the interaction between two spherical Janus particles.
In \cite{SharifiMood:2015}, both hydrodynamics (the Stokes equation) and a
concentration field (the Laplace equation) are solved.
In contrast to \cite{Papavassiliou:2016}, axisymmetric motion and
rotation were not assumed. 
Nevertheless, as the results of infinite series of coefficients were evaluated
numerically a direct comparison is also not possible here.

We treat the near-field interactions in a similar way to  \cite{ishikawa:2006,Ishikawa:2008a}, however  
while they compute passive and active resistance matrices using numerical
tables, we obtain explicit formulas for the translational 
and angular velocity as \NY{eqs.~}(\ref{trans.vel}) and (\ref{ang.vel}).
This allows us to perform simulations of this problem in 
an analogous way to active Brownian Particle simulations.
We note also that while the force in the direction  perpendicular  to the centreline and torque were given 
 in \cite{ishikawa:2006}, an explicit expression of the force in
 the parallel direction was not provided in that paper.
In \cite{ishikawa:2006}, the active force in the perpendicular direction to the centreline between two
spheres was obtained as $F_{x,1}^{(a)} = -\pi \eta R d u_s \log \epsilon$,
where $d u_s$ is ``the difference between the squirming velocities of
the two squirmers'' and is expressed in our notation as $d u_s =
\sum_{l,m=\pm1} m \mathcal{N}_{l|m|} V_{lm}$.
This result is exactly same as our result here, given by the first term in
\NY{eq.}(\ref{nonaxis.Fx1}) once we make the assumption of uniaxial slip
 velocity using \NY{eq.~}(\ref{uniaxial.slip.corrsp}).
Similarly, the torque was obtained as $T_{y,1}^{(a)} = -\pi \eta R d u_s
\log \epsilon$ when the sphere 1(2) is moving in the $x$($-x$)
direction under the configuration shown in fig.~\ref{fig.sec1.8problems}
(a).
This implies $W_{lm}=0$ in our notation, and then in the uniaxial case, \NY{eq.~}(\ref{nonaxis.Ty1}) gives
exactly the same result using the properties of the spherical harmonics
and the associated Legendre polynomials.
We note that the analytical calculation in \cite{ishikawa:2006} is based on calculations of the force 
and torque when one sphere has a slip velocity and the other
sphere is inert. Explicit formulas are given in \cite{ishikawa:2006} only  for the force and torque acting on the
squirming sphere, and not for inert one. Here we find that in
 order to evaluate the general form of the torque acting on both
spheres when they are both squirming, the torque acting on the inert
sphere is necessary because (g) in fig.~\ref{fig.sec1.8problems} has the
same singularity as (c). It is not clear how  this issue was treated in \cite{ishikawa:2006}.
  We note also that although the we obtain identical expressions for the perpendicular forces and torques to \cite{ishikawa:2006}  for two spheres of the same size in the limit  $\epsilon \rar 0$,
  we notice that the force and torque we obtain for squirmers of different sizes  ($\alpha \neq 1$ in \cite{ishikawa:2006}) disagree with  \cite{ishikawa:2006} for the same system (leading to different behaviour, e.g. , for the interaction
  between a squirmer and a wall).
  We are not aware of the comparison between numerical and  analytical
  results in such cases.

The force in the direction 
parallel  to the line between centres, 
 is estimated by \cite{ishikawa:2006} to be  proportional
 to $\log \epsilon$ although they did not provide an explicit expression. 
 The coefficient of $\log \epsilon$ in the force along the
parallel direction to the centreline between two spheres, and the
$\mathcal{O}(1)$ terms of the force in the perpendicular direction and
the torque are evaluated by comparison with the numerical results of
Boundary Element Method.
The force in the parallel direction was estimated as $F_{z,1}/(6\pi \eta R) = -2.4 \log \epsilon$~\cite{ishikawa:2006} while
we obtain from our leading order analysis that
\begin{align}
 \frac{F_{z,1}}{6\pi \eta R} &= -2 \log \epsilon
\end{align}
from \NY{eq.~}(\ref{axisym.Fz1}).
We also note the our expression becomes same as the analytical result of the singular
term in  \cite{Wuerger:2016} for the parallel interaction between a
squirmer and a wall, that is $F_{z}$ in our notation.

We did not compute the $\mathcal{O}(1)$ terms and instead use the values
of an isolated squirmer as was done in \cite{Swan:2011} for
a Stokesian Dynamics simulation.
This is not a bad approximation particularly for a near
contact since the dominant contribution arises from the singular terms.
In fact, the torque between the two squirmers moving in the same
direction parallel to their centreline is numerically shown to be $\mathcal{O}(1)$
times the torque for the an isolated rotating sphere \cite{ishikawa:2006}.
Therefore, we estimate that our method is of comparable  accuracy to 
other methods for the near-field interactions.

Finally, as briefly mentioned above, we  note that 
the study of \cite{ishikawa:2006} assumed
uniaxial slip flow on the surface of a squirmer.
Note that this does not mean axisymmetry of the motion and rotation was
assumed.
In our approach, we do not need to assume axisymmetric slip flow although we have made the assumption to simplify our
numerical simulations.
For example, the dipolar force described by $v_{2,m}$ could  have a different
axis to the  swimming direction described by $v_{1,m}$.
In addition, the dipolar force is not necessarily uniaxial but can be
biaxial.
In that case, the dipolar force $v_{2,m}$ is expressed by two amplitude
and three angles.
As we have seen in \NY{eq.~}(\ref{v2.5d.vector}), the uniaxial flow is expressed
by one amplitude (denoted by $v_2$) and two angles (denoted by $\alpha$
and $\beta$).
The biaxial flow has another angle in the plane perpendicular the axis
expressed by $\alpha$ and $\beta$.

So in summary,  we typically find that for all the quantities calculated in \cite{ishikawa:2006}, we find agreement between our results and those in\cite{ishikawa:2006}, however in order to accurately describe the interactions between arbitrarily oriented pairs of squirmers, we need to calculate additional quantities not presented in \cite{ishikawa:2006}. Therefore, our method can be viewed as the minimal mathematically complete enumeration of interactions between near contact squirmers, which entails a   generalization of the analytical results in \cite{ishikawa:2006}.

Stokesian Dynamics was used in \cite{Swan:2011}  to simulate collections of swimmers.
However, the model used in the work is not a squirmer, which has slip
boundary condition \NY{eq.~}(\ref{janus.intro.slip}), but rather another class of
swimmer; in which the moments of force acting on the particle are imposed.
With this model, it is not necessary to compute the active resistance
matrix but the dynamical equation requires only the passive resistance
matrix, which is known in literature.

We consider only pairwise hydrodynamic interactions, and therefore the
many-body effects we observe arise from the superposition of pairwise interactions.
Our method does not capture many-body interactions that are
not described by the sum of pair-wise interactions.
An improvement of our model could be obtained 
 by computing the lubrication interactions between higher moments of the local velocity
fields associated with each squirmer; for example, including a stresslet contribution to the resistance matrices~ \cite{brady:1988,ishikawa:2006}.
A recent work using pairwise interactions computes
 the grand mobility and resistance matrices~\cite{Singh:2016} using tensorial spherical harmonics and the Galerkin
method for the integral representation of the Stokes flow.
This is similar to the incorporation of higher order terms of a (far field) multipole moment expansion in passive
colloidal suspensions, such as performed in \cite{brady:1988}(up to Stresslet),
\cite{Ichiki:2002} (arbitrary order in a real space), and
\cite{Mazur:1982,Ladd:1990} (arbitrary order in a Fourier space).
 However, in passive suspensions, it is known that the far field multipole moment
 expansion, even with higher order terms, converges slowly or does not
 converge at all in the limit (close proximity of particles) where lubrication interactions dominate.
 This was the motivation for adding the lubrication correction between all passive 
 pairs in \cite{brady:1988,Ladd:1990} using exact limiting
 formulas.
  A possible modification of the approach in~\cite{Singh:2016}, is to include the lubrication corrections by hand
 using the exact limiting formula of the active
 resistance matrix computed in this work and in \cite{ishikawa:2006}, or
 , as proposed in \cite{Singh:2016}, by numerically solving the integral
 equations of the boundary element method.
 The former method gives accurate formulas in the limit of close contact,
 but extending it to higher order moments might not be easy as we are not aware of systematic studies on lubrication forces as
 a function of higher moments (higher than the stresslet contribution).
 The latter method is straightforward, but is cumbersome as it gives  a numerical
 table, which can be called in a many-body simulation.
 We stress, however that simply including the higher order far field moments themselves does not capture the
 lubrication forces accurately because they make singular contributions to the dynamics.
 Note that our formula may be expressed also by using the Cartesian
 tensors as demonstrated in Appendix.~\ref{sec.Cartesian}

We note that standard Lattice Boltzmann simulations and Multi-Particle
Collision Dynamics simulations are not designed to compute near-field
interactions because of the limitation of a small mesh size and absence
of ideal-gas particles to transport momentum when two particles are too
close to one another.
For example, in \cite{Alarcon:2013}, the size of a particle, $R=2.3$ or $R =4.7$ in
the unit of the mesh size was used and the time scale is normalised by
$R/u_0$.
Therefore, the interaction is not accurate  when $\epsilon \lesssim 0.2$.
Similarly, in \cite{Zoettl:2014}, the particle size is chosen as $R=3$
in the unit of the mesh in the collision step.
Therefore, the simulation has similar accuracy of the near-field
interactions with that in \cite{Alarcon:2013}. Our approach, in contrast is particularly good at short-range interactions when $\epsilon \ll 1$ because lubrication forces
arise only near the contact points between two particles.

Our method, however, fails to describe intermediate-distance
hydrodynamic interactions properly in contrast with the methods such as
Multi-Particle Collision Dynamics and Lattice-Boltzmann Method.
This may be improved by computing the higher-order terms in the matched
asymptotic expansion~\cite{Jeffrey:1984}.
\\

\section{Collective behaviour}
\label{sec.collective}

Using the explicit expressions for the hydrodynamic interactions outlined above, we performed numerical simulations with $N$ identical particles of radius
$R$ 
with periodic boundary conditions.
We use up to $N=8192$ particles for computations including the near and far field interaction,
and use up to $N=32768$ particles for the computations only with near-field
interactions.
Defining, the average distance between two particles, $\xi = R \sqrt{\pi/\rho_0}$,
we vary $\xi$ from $\xi \simeq 2.65$ to $\xi \simeq 5.30$.
We set $v_1 =1$ for all swimmers and thus $u_0 \simeq 0.32$.
The size of a particle is chosen to be of unit length, thus we set $R=1$.
The time scale is normalised by the time for an isolated squirmer to
move a half of its body length, that is, $\tau_0 = R/u_0$.
There is a time scale associated with  collisions, $\tau_m = \xi/u_0$.
 We vary the time scale from $\tau_m \simeq 8$ to $\tau_m \simeq 17$.
We consider motion restricted to 2D {\em but} interacting via 3D hydrodynamic interactions as well as fully 2D systems.
We neglect the modes with $l \ge 3$ for three-dimensional systems and $m
\ge 3$ for two-dimensional systems.
We note that pushers ($v_2>0$) and pullers ($v_2<0$) having the same $|v_2|$ with 3D hydrodynamics are not identical when constrained to move in 2D. This is  
because, defining orientations with respect to the swimming direction, the interactions at the front and the back are stronger than those at the sides.
As a result, pullers, on average attract nearby objects.
Pushers and pullers with 2D hydrodynamics are only equivalent when quadrupoles are absent,
but, in general, as long as they have finite quadrupoles, they will not be identical.

\begin{figure}[htbp]
 \begin{center}
    \resizebox{0.5\textwidth}{!}{%
  \includegraphics{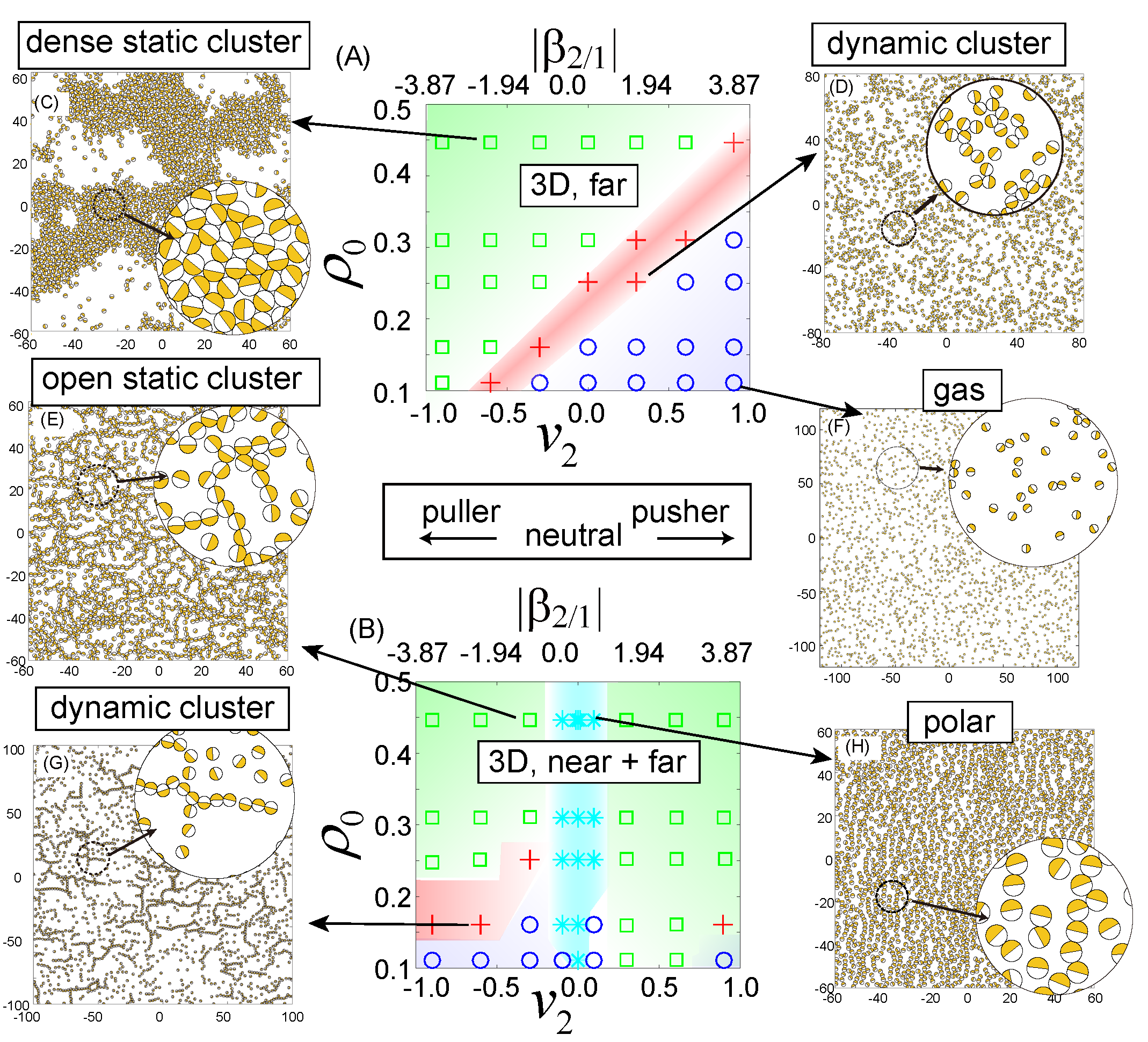}
}
 \caption{
 (Colour Online) The state diagram of squirmers with density ($\rho_0$) and dipolar strength ($v_2$) with (A) far field
 hydrodynamic interactions and (B) both far field and near field in 3D
 systems with $N=2048$ particles.
  Snap shots of (C) \NYY{the dense static cluster state ($\rho_0=0.447$, $v_2=-0.6$)},
 (D) the dynamic cluster state ($\rho_0=0.251$,$v_2=0.3$), and (F) the disordered state ($\rho_0=0.112$,$v_2=0.9$) for a far-field-only
 system, and (E) \NYY{the open static cluster state ($\rho_0=0.447$, $v_2=-0.3$)}, (G) the dynamic cluster state \NYY{($\rho_0=0.161$,$v_2=-0.3$)}, and (H) the
 polar state \NYY{($\rho_0=0.447$, $v_2=0.1$)} for a near+far system.
 \NYY{The inset of each figure is a close view of position and
 orientation of particles.}
 \label{fig.Phase.Diagram}
}
 \end{center}
\end{figure}

Our numerical results are summarised by the phase behaviour shown in
fig. \ref{fig.Phase.Diagram} as a function of density $\rho_0= \pi R^2
N/L^2$ and the force dipole strength $v_2$.
The phase diagram has been generated for systems with $N=2048$ particles.
In studying the phase behaviour, we have emphasised the dependence on
the sign of the force dipole  ($v_2 \neq 0$) and contrasted them to neutral swimmers with force quadrupole and no force dipole ($v_2=0$). 
 We find significant differences between the hydrodynamic interactions with
 and without near-field effects.
 The general structure of the phase diagrams is that 
at low densities, the system has a disordered  `gas'  state and at  higher
densities, there is the emergence of stable clusters except for neutral
swimmers ($v_2=0$).
We note that the behaviour of neutral swimmers with {\em far-field} interactions only are similar to those of
 ABPs since there is no rotation induced by collisions.
 Upon including near-field effects, neutral swimmers spontaneously develop polar order.
 Finally, we find that stable clusters are suppressed by the near-field
 interactions, leading to dynamical clusters of finite size that exchange particles with bulk.
 The phase boundaries are qualitative; we set a specific values of
 threshold to distinguish between two phases. 
 Nevertheless, as we will see in sect.~\ref{sec.polar}, the phase
 boundary between the polar state and other states, for a fixed density of swimmers, are independent of threshold and easily 
 identified.
 We also note in sect.~\ref{sec.clusters} we will see there is no clear
 definition of phases between dynamic and static clusters.

\subsection{Static and dynamical clusters}
\label{sec.clusters}

\tlll{As the density of swimmers is increased, they collide more often with one another and possibly form aggregates (clusters) of many swimmers.
We identify two classes of  clusters which we denote as : static (SC) or
dynamic (DC).
Static clusters grow irreversibly until the majority of the swimmers are in one large cluster while dynamic clusters exchange particles with the bulk and remain of  finite size.
The analysis is motivated by 
the appearance of dynamical clusters of active particles in recent
experiments \cite{Palacci:2013,Buttinoni:2013}.
\tlll{In two-dimensional (2D) ABP
systems~\cite{Fily:2012,Buttinoni:2013,Redner:2013,peruani:2006} and for
squirmers confined between walls \cite{Zoettl:2014,Blaschke:2016},
macroscopic phase separation 
has been observed associated with a clustered state.
 A major difference however is finite size clusters in
 experiments~\cite{Palacci:2013,Buttinoni:2013} while the infinite
 cluster is formed in the  macroscopically phase separated state.
 In ABPs, rotational relaxation time is assumed to independent of a collision.}
 Since hydrodynamic interactions result in rotation of interacting
 particles, this assumption is not valid for squirmers.
In fact, recent simulations have shown that clusters are absent both in 
2D squirmer suspensions and in a 2D squirmer monolayer embedded in a 3D fluid~\cite{Matas-Navarro:2014,Saffman:1975}.}

\begin{figure}[htbp]
   \begin{center}
    \resizebox{0.5\textwidth}{!}{%
  \includegraphics{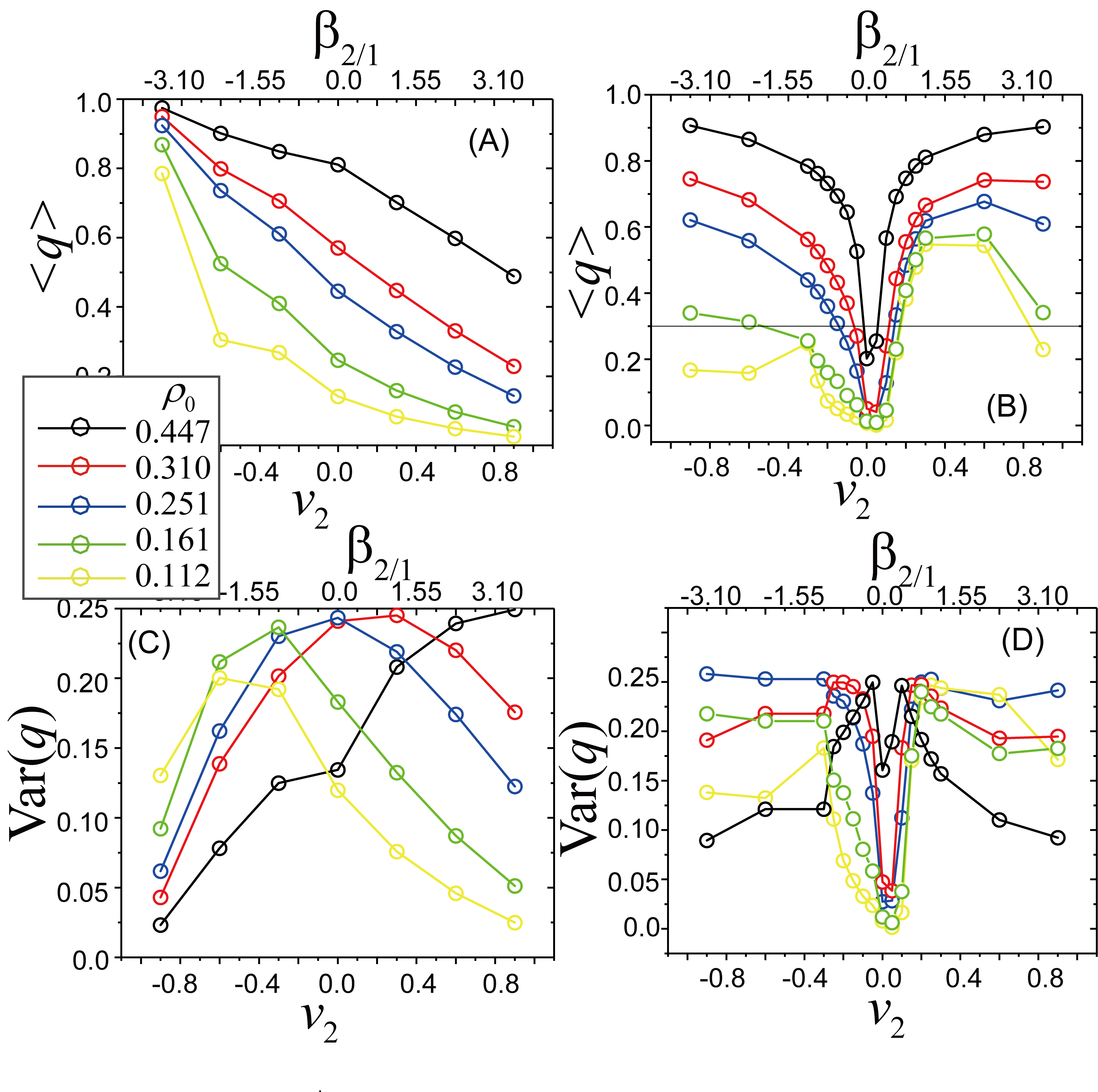}
}
 \caption{
 (Colour Online) (A, B) The cluster ratio, $\mean{q}$ and (C, D) cluster
 fluctuation, $\mean{q^2}$ with (B,D) and without (A,C) near-field interactions as a function
 of dipolar strength, $v_2$.
\label{fig.analysis}
}
 \end{center}
\end{figure}

Cluster formation (see fig.~\ref{fig.analysis}(A)-(D)) is analysed as follows.
Two swimmers with centres separated by less than $r_c = 2R
+ 0.01$ are in the same cluster. The $i$th particle is labeled as $q_i(t) = 1(0)$ when it is (not) in a
cluster.
  We define the cluster ratio as $q(t) = (1/N)\sum_{i=1}^{N} q_i(t)$.
Clustered phases are defined as those with cluster ratio $\mean{q} > 0.3$ where the average is taken over particles
  and time in the steady state.
This threshold is chosen from the value of $\mean{q}$ at the maximum
  value of $\mean{q^2}$ when $v_2$ is varied under different densities.
  In contrast to the boundary of polar order, the boundary between gas and cluster
  phases is not a sharp transition but rather a smooth cross-over. 

 We discriminate between the clusters of the two types: static (SC) or 
  dynamic (DC), by analysing the dynamics of particles into and out of the clusters. The variance of the clustered state $\mean{q^2}$ along the trajectory of
  each particle, averaged over all particles in the steady state
  shows how often the particles are exchanged between clustered and
  non-clustered states. By modeling $q_i(t)$ as a Telegraph process~\cite{gardiner:1985},  these quantities are associated with on the rate of each particle
  joining a cluster ($k_{\rm on}$) and leaving it ($k_{\rm off}$).
These rates introduce time scales: the residence time inside a cluster
  $\tau_{\rm res} = k_{\rm off}^{-1}$, and the time scale to jump from one
  cluster to another,  $\tau_{\rm out} = k_{\rm on}^{-1}$, in addition to the time scale of a
  collision, $\tau_m$.
  For the Telegraph process, the mean value and its variance are given
  by
  \begin{align}
   \mean{q}
   &=
   \frac{  k_{\rm on}}{k_{\rm on} + k_{\rm off}}
   \\
      \mean{q^2}
   &=
   \frac{  k_{\rm on} k_{\rm off}}{ (k_{\rm on} + k_{\rm off})^2}
  \end{align}
Therefore  the ratio between the two rates is given by
\begin{align}
 \frac{k_{\rm on}}{k_{\rm off}}
 &=
 \frac{ \mean{q}^2}{\mean{q^2}}
\end{align}
  \cite{gardiner:1985}.
  The time correlation function of this process is expressed by
  \begin{align}
   \mean{q(0) q(t)} -  \mean{q}^2
   &=
   \mean{q^2}
   e^{ - \left(
k_{\rm on} + k_{\rm off}
   \right)
   t}
  \end{align}
  From the correlation time ($1/(k_{\rm on} + k_{\rm off})$) obtained by fitting this correlation function, we may evaluate $k_{\rm on}$ and $k_{\rm off}$
  (equivalently,  $\tau_{\rm out}$ and $\tau_{\rm res}$) separately.
 Hence we can define the clustered states (both SC and DC) by
  $\tau_{\rm out} > \tau_m$ in which clusters do not form even after
  collision.
  Our threshold above in terms of $\mean{q}$ gives the same results. 
 At higher densities, the particles stay in 
  clusters for longer periods, namely $k_{\rm on}/k_{\rm off} > 1$ and
  $\tau_{\rm res} > \tau_{\rm out}$.
  This state is classified as the static cluster (SC).
  The phase boundaries between SC and DC and between DC and gas may
  shift under the change of the
  threshold of $k_{\rm on}/k_{\rm off}$ although we confirmed the
  qualitative behaviour does not depend on it.

\subsection{Polar state}
\label{sec.polar}

In this section, we discuss the emergence of polar order and its stability.
Polar order has already been reported for neutral squirmers  with no force dipole in monolayer in 3D (i.e. 3D interactions, but  motion in 2D)
\cite{Ishikawa:2008a} and in 3D \cite{Evans:2011,Alarcon:2013,Oyama:2016}.
It has been suggested  \cite{Ishikawa:2008a} that near-field effects {\it
enhance} the polar state although a weakly polar state also appears 
using the far-field approximation~\cite{Ishikawa:2008a}.
This is in contrast with well established continuum arguments which demonstrate that the polar state is generically unstable for wet
active systems~\cite{Marchetti:2013,Toner2005,ramaswamy:2010}.
This has raised  the interesting possibility of other 
continuum limits in these systems. Due to the expense of computational treatment of hydrodynamics however, the previous results~\cite{Ishikawa:2008a,Evans:2011,Alarcon:2013,Oyama:2016} are limited to relatively small numbers of
particles, and questions remain about whether a macroscopic continuum limit had truly been achieved. Achieving this limit is one of the goals of 
this work.

\begin{figure*}[htbp]
   \begin{center}
    \resizebox{0.8\textwidth}{!}{%
  \includegraphics{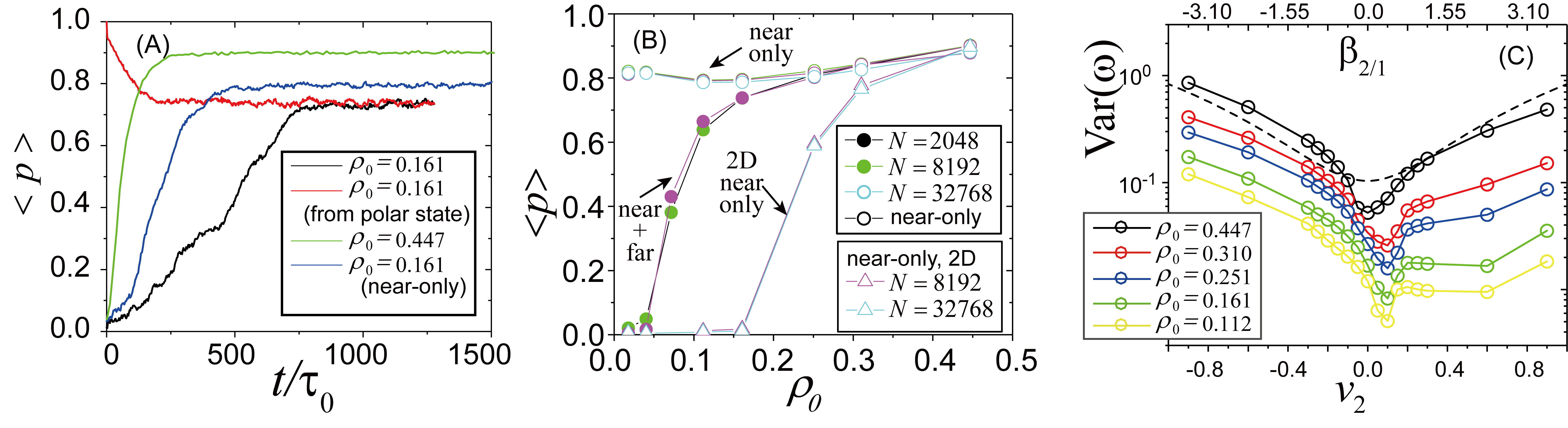}
}
 \caption{
 (Colour Online)
 \NYY{(A) Time evolution of $\mean{p}$ for neutral swimmers
 at $\rho_0=0.16$ ($\xi = 4.43$) shown in black (the initial condition
 is disordered state) and red (ordered state) and $\rho_0=0.447$ ($\xi =
 2.65$) shown in green.
 The polar order of the near-field-only system is shown in blue.
 }
(B) the mean polarity, $\mean{p}$ for neutral swimmers
 $v_2=0$ vs density, $\rho_0$.
 \NYY{
The solid (open) circles correspond to the simulations with (without)
 the far-field interaction in 3D systems.
 The open triangles correspond to the results of the near-field
 interaction in 2D systems. 
 }
 Both include the near field interaction.
 (C) rotational fluctuation, $\mean{\omega^2}$ vs dipolar strength,
 $v_2$.
 The dashed line shows $\mean{\omega^2} \sim v_2^2$ for the guide for the eyes.
\label{fig.analysis2}
}
 \end{center}
\end{figure*}

\begin{figure}[htbp]
 \begin{center}
    \resizebox{0.5\textwidth}{!}{%
  \includegraphics{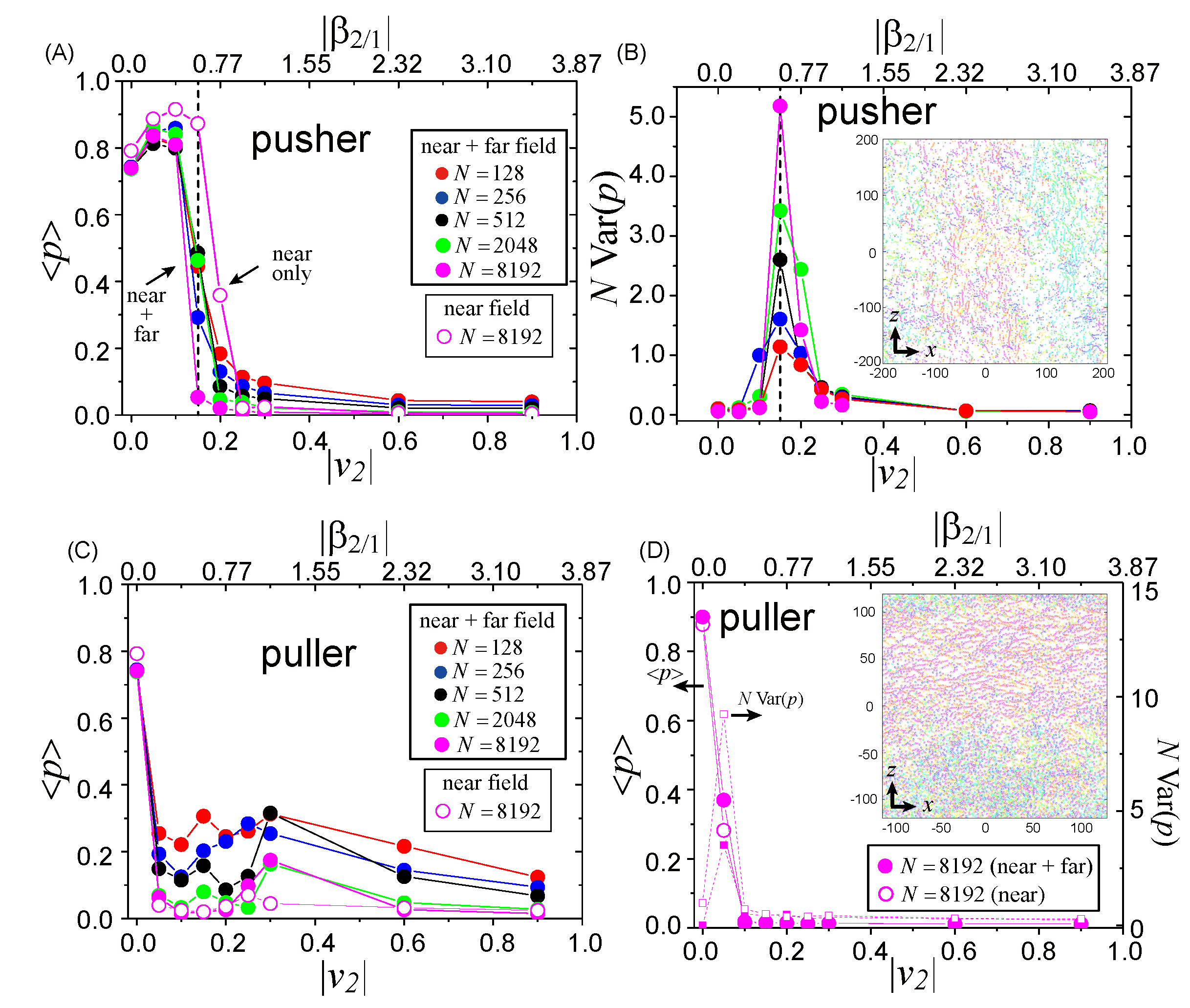}
}
 \caption{
  (Colour Online)
  The mean polarity as a function of the dipolar slip
  flow $v_2$ on a squirmer at $\rho_0=0.16$ ($\xi = 4.43$).
 The corresponding values of the squirmer index $\beta_{2/1}$ is shown
 at the upper axis.
 The polar order $\mean{p}$ for pushers (A) and pullers (C).
 The near and far field systems are shown by the filled circles while the
 systems only with near field interaction is shown by the open circles. 
  (B) Fluctuation for pushers.
  \NYYY{The inset shows a snapshot at $v_2=0.15$ for $N=8192$.
 The colour shows orientation angle of each particle from $z$ axis (red:
 $0,2\pi$, green:$2\pi/3$, and blue:$4\pi/3$).
  }
 (D) \tlll{The mean polarity and 
 fluctuations about the mean  for pullers} at $\rho_0=0.447$ ($\xi = 2.65$).
 Both the results of the near+far (solid circle) and the
  near-field-only (open circle) systems are shown.
  \NYYY{Fluctuation for pullers is shown by squares.
   The inset shows a snapshot at $v_2=-0.05$ for $N=8192$.}
\label{fig.v2}
}
\end{center}
\end{figure}

For neutral (quadrupolar) squirmers and squirmers with small dipoles, $\left |v_2 \right | \ll 1$, the formation of clusters is significantly suppressed and 
they rather self-organise into a polar state in which the swimmers align
their orientations and swim in the same direction (fig.~\ref{fig.analysis2}(A)). 
 The polar state is robust to changes in the density
  but for very low density, polar order vanishes.
  The development of polar order is independent of initial condition. Furthermore,
 the amount of polar order does not change much at  higher densities,
 but the relaxation time does become shorter.
Since
 the far-field interaction does not cause reorientation, polar order is purely driven by near-field
 effects.
 In fact, upon screening the far-field interaction which may happen
  at high density\cite{Ball:1997}, 
  we obtain almost identical 
  polar ordered states in figs.~\ref{fig.analysis2}(A) and (B) (also in
  fig.~\ref{fig.v2}(D)).
 On introducing rotational noise, the disorder-polar transition is still present but 
  occurs at a higher density.

In the phase diagram (see fig.~\ref{fig.Phase.Diagram}), the polar state
corresponds to $\mean{p} =(1/N)| \sum_{i} {\bf p}^{(i)}| > 0.7$.
This value is chosen from the value of $\mean{p}$ above the transition
from the gas state to the polar state induced either by pusher dipole
force or the rotational noise (see fig.~\ref{fig.v2}(A,B) and
fig.~\ref{fig.noise}).
We emphasize that this  stringent criterion means that the polar regions of the phase (state) diagram correspond to strongly ordered globally polar states.


  The polar state is present for small but nonzero $v_2$, but as $v_2$
  is increased, polar order vanishes \NYY{as shown in
  figs.~\ref{fig.v2}(A), (C)
  and (D).}
  This is due to large fluctuations of the angular velocity (see fig.~\ref{fig.analysis2}(C)),
  $\mean{\omega^2} \sim v_2^2$ from \NY{eq.~}(\ref{ang.vel}).
%
 For pushers ($v_2 >0$), polar order disappears around $v_2 \simeq 0.15$
 ($\beta_{2/1} \simeq 0.58$).
 The transition from the polar-orderd state to a non-polar state is accompanied by divergence of fluctuations of the polar
 order as shown in fig.~\ref{fig.v2}(B).
 On the other hand, pullers  ($v_2  < 0$)  behave somewhat differently as one increases 
 $|v_2|$.
 First, for pullers, the amount of polar order fluctuates more than for pushers, and
 accordingly the mean polar order parameter $\mean{p}$ becomes noisy.
 Second, we do not observe a distinct peak in the fluctuations in polarity between the polar
 order and the disorder state.
Nevertheless, when the density is high ($\rho_0=0.447$ and $\xi =
 2.65$), there is clearer divergence of fluctuation as shown in
 fig.~\ref{fig.v2}(D).
 At the transition point, there is phase separation between polar order
 and disorder.
 For large enough $|v_2|$,  polar order completely vanishes for both
 types of swimmers.
 \NYY{
 The polar order still present even when the far-field interactions are
 switched off.
The critical dipolar force is slightly larger for such systems for pushers, but the
 polar order is dependent on the dipolar force $v_2$ similar to the near
 and far field interactions (fig.~\ref{fig.v2}(A)).
 For pullers, the near field interactions also suppress polar order
 while the noisy behaviours discussed above vanish in this case
 (fig.~\ref{fig.v2}(C)) suggesting that it arises from the far-field interactions.
 }

\begin{figure}[htbp]
 \begin{center}
    \resizebox{0.5\textwidth}{!}{%
  \includegraphics{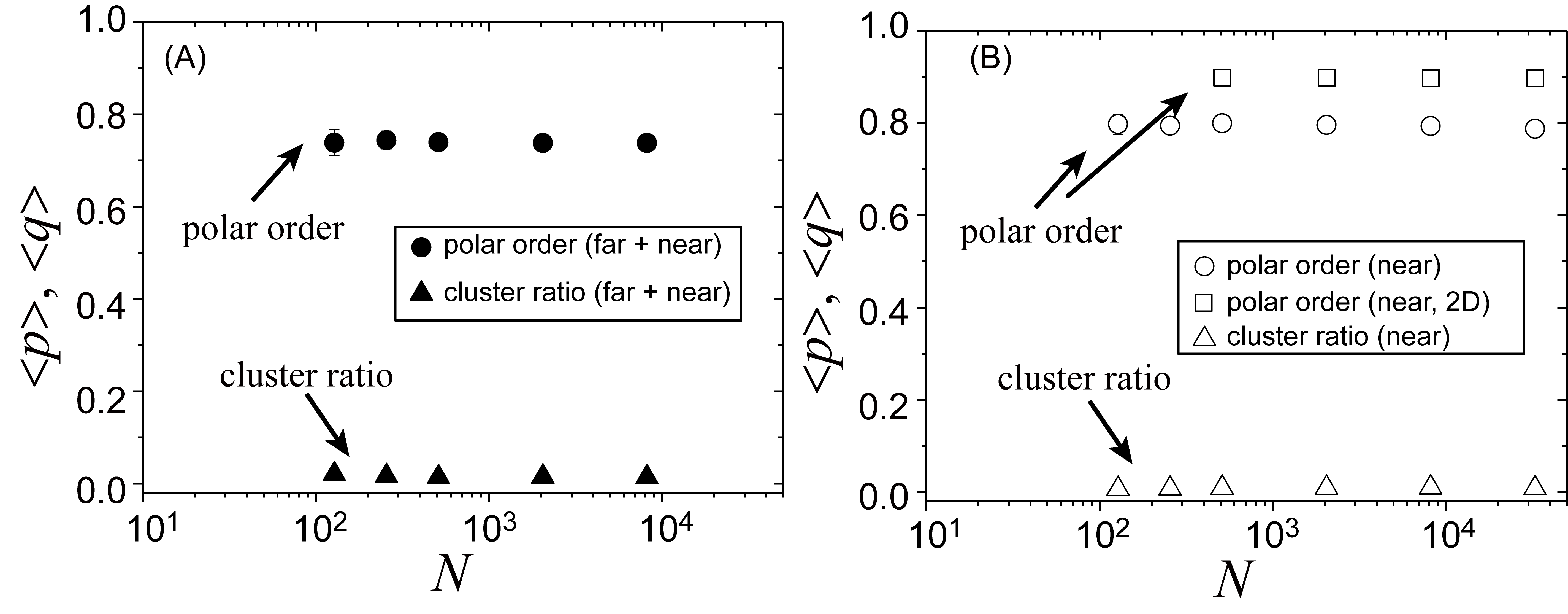}
}
 \caption{
 (Colour Online)
\NYY{
 The system size dependence of the mean polar order
 $\mean{p}$ and the mean cluster ratio $\mean{q}$ at $\rho_0=0.16$
 ($\xi = 4.43$) of the neutral swimmer $v_2=0$ for (A) far- and
 near-field systems and for (B) near-field-only systems.
 in (B), the results of two-dimensional neutral squirmers for
 $\rho_0=0.447$ ($\xi = 2.65$) are also shown in the open triangles. 
 }
\label{fig.Ndependence}
}
\end{center}
\end{figure}

The polar order is stable even for very large numbers of particles.
In fig.~\ref{fig.Ndependence},  polar order is shown as a function of
the number of particles, $N$ up to $N=8192$ for the near+far field system
and $N=32768$ for the near-field-only system.
Polar order survives even for large system sizes, and therefore we conclude that 
the system is asymptotically in a state with macroscopic global polar order.
The simulations without far-field interactions give slightly larger
stronger polar order for the same density.
Nevertheless, the systems with and without far-field interactions show
the same behaviour.
The mean cluster ratio of the polar phase is nearly zero throughout the
range of system sizes, indicating that there is no clustering associated with  these polar states.

\subsection{\tll{Orientational fluctuations}}

\begin{figure*}[htbp]
 \begin{center}
    \resizebox{0.8\textwidth}{!}{%
  \includegraphics{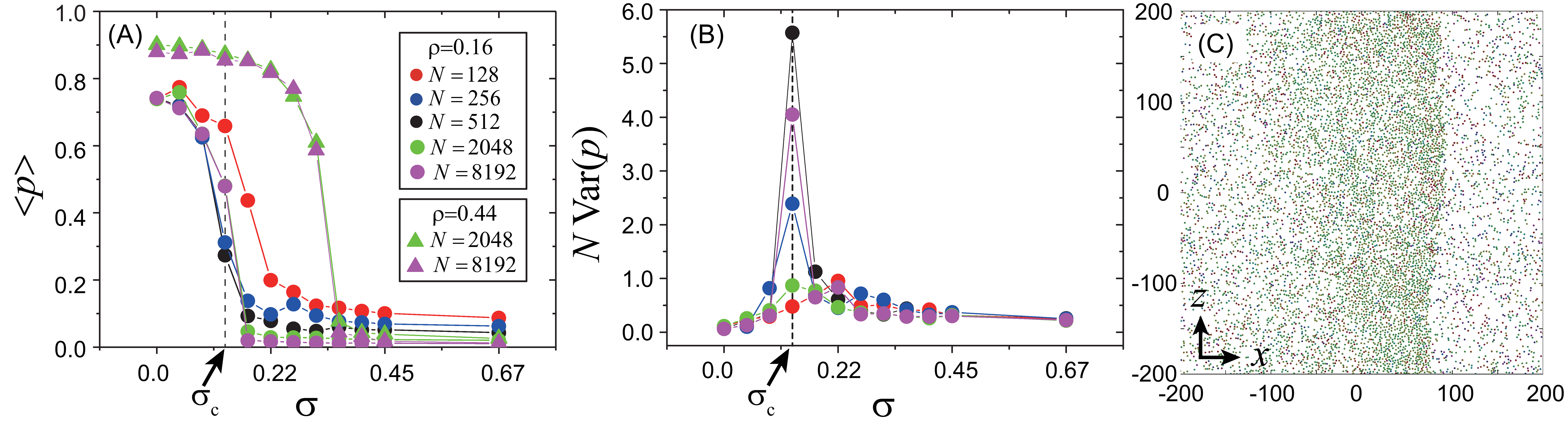}
}
  \caption{
  (Colour Online)
\tlll{
  (A) The mean polarity as a function of orientational noise amplitude at $\rho_0=0.16$ ($\xi = 4.43$, closed circles) and at
  $\rho_0=0.447$ ($\xi = 2.65$, closed triangles) for the neutral swimmers
  ($v_2=0$ and $v_1=1.0$ in \NY{eq.~}(\ref{squirmer.u0})).
}
  (B) The variance of the polarity. The dashed line
 corresponds to the critical point where $\sigma_c = 0.13$ \NY{for $\rho_0=0.16$}.
 (C) The polar band state near the critical amplitude of noise \NYYY{for
  $N=8192$} \NY{for $\rho_0=0.16$}.
 The colour shows orientation angle of each particle from $z$ axis (red:
 $0,2\pi$, green:$2\pi/3$, and blue:$4\pi/3$).
\label{fig.noise}
}
\end{center}
\end{figure*}

\tlll{
To examine the stability of polar order to fluctuations, we add
uniformly distributed
white noise, customarily used in the Vicsek model \cite{chate:2008}, to the angular velocity of each swimmer.
To be precise, position and orientation are updated by
\NY{eqs.~}(\ref{janus.dotr.u}) and (\ref{janus.dotp.omega}) as before. 
However the
angular velocity defined \NY{eq.~}(\ref{ang.vel}) now has an additional  fluctuating component and  becomes
 \begin{align}
  {\bm \omega}^{(i)} 
&=
 \sum_{j \neq i, l,m}
\omega_{lm}^{(j)}
 {\bm \Phi}_{lm}^{(ji)}
  + {\bm \omega}^{(i)}_f(t)
 \label{ang.vel.noise}
 \\
 \mean{{\bm \omega}^{(i)}_f (t)}&=0 \; , \; 
 \mean{{\bm \omega}^{(i)}_f (t) \cdot {\bm \omega}^{(i)}_f (t')}
 =
 \frac{\sigma^2}{3}
 \delta_{ij}
 \delta(t - t')
 \label{eq:rot_noise}\end{align}
 for a system with fixed density $\rho_0$. We emphasize that no noise is added to the translational velocity,
\NY{eq.~}(\ref{trans.vel}) and any additional fluctuations in position arise purely from rotational noise, eq.~(\ref{eq:rot_noise}).
}
 \NY{
In our quasi-2D system, rotation induced by the noise occurs only in the plane. 
 }
The amount of polar order as a function of the amplitude $\sigma^2$ of the rotational noise is shown in
fig.~\ref{fig.noise}.
 Around $\sigma = \sigma_c \simeq 0.13$, we observe  a  phase transition
 between a polar state and a gas state \NY{at $\rho_0=0.16$}.
The polar state
 is present for lower noise amplitudes while the gas state is stable  at
 higher noise amplitudes. 
 \tlll{
 The critical noise amplitude, $\sigma_c$ is density dependent. 
At higher densities of squirmers, ($\rho_0=0.44$),  the transition from the polar state to the gas state occurs at higher noise amplitudes.
For both densities studied ($\rho_0=0.16,0.44$), the polar ordered state has a uniform density with  no observable micro-phase separation.
 }
The fluctuations of the polarity about its mean value diverges as the amplitude of noise
approaches its critical value $\sigma_c$.
Interestingly, a polar band appears near the critical point, in which
the low density gas state and high density polar state coexist.
This behaviour is similar to the gas-polar phase transition of
the Vicsek-type models (fig.~\ref{fig.noise}(C)).

\tlll{
We characterise the rotational noise by a rotational diffusion
coefficient $D_r$, or equivalently a rotational diffusion time,
$\tau_r$
\begin{align}
 \tau_r
 &=
 \frac{1}{2 D_r}
 = \frac{3}{\sigma^2}
\end{align}
The rotational diffusion time at the critical noise is $\tau_r =
3/\sigma_c^2 \simeq 178$.
We can define a  rotational P\'{e}clet number, 
\begin{align}
 {\rm Pe}
 &=
 \frac{u_0 \tau_r}{R}
 \label{peclet}
\end{align}
which compares the self-propulsion speed and rotational diffusion.
At the critical noise, ${\rm Pe} \simeq 60$ for $\rho_0=0.16$ and ${\rm
Pe} \simeq 10$ for $\rho_0=0.44$.
Although the rotational P\'{e}clet number for $\sigma \ll \sigma_c$ for
the higher density
is in the region where macroscopic phase separation
 has been observed for active brownian particles~\cite{Stenhammar:2014}, in our system we find a uniform density of
squirmers except in the vicinity of the transition from the polar to gas phase where the noise amplitude has its critical value, $\sigma_c$(Fig.~{fig.noise}(C)).
Note, however, the rotational P\'{e}clet number \NY{eq.~}(\ref{peclet}) is only 
based on the noise added to the system, and does not take account of additional rotational fluctuations generated by collisions, which occur even in 
deterministic squimers  (see
fig.~\ref{fig.analysis2}(C)).
This is consistent with previous results that suggest the suppression of 
macroscopic phase separation
 in squirmer suspensions~\cite{Matas-Navarro:2014}.
}

\subsection{Collective behaviours of two-dimensional squirmers}
\label{sec.2D}

\begin{figure}[htbp]
 \begin{center}
    \resizebox{0.5\textwidth}{!}{%
  \includegraphics{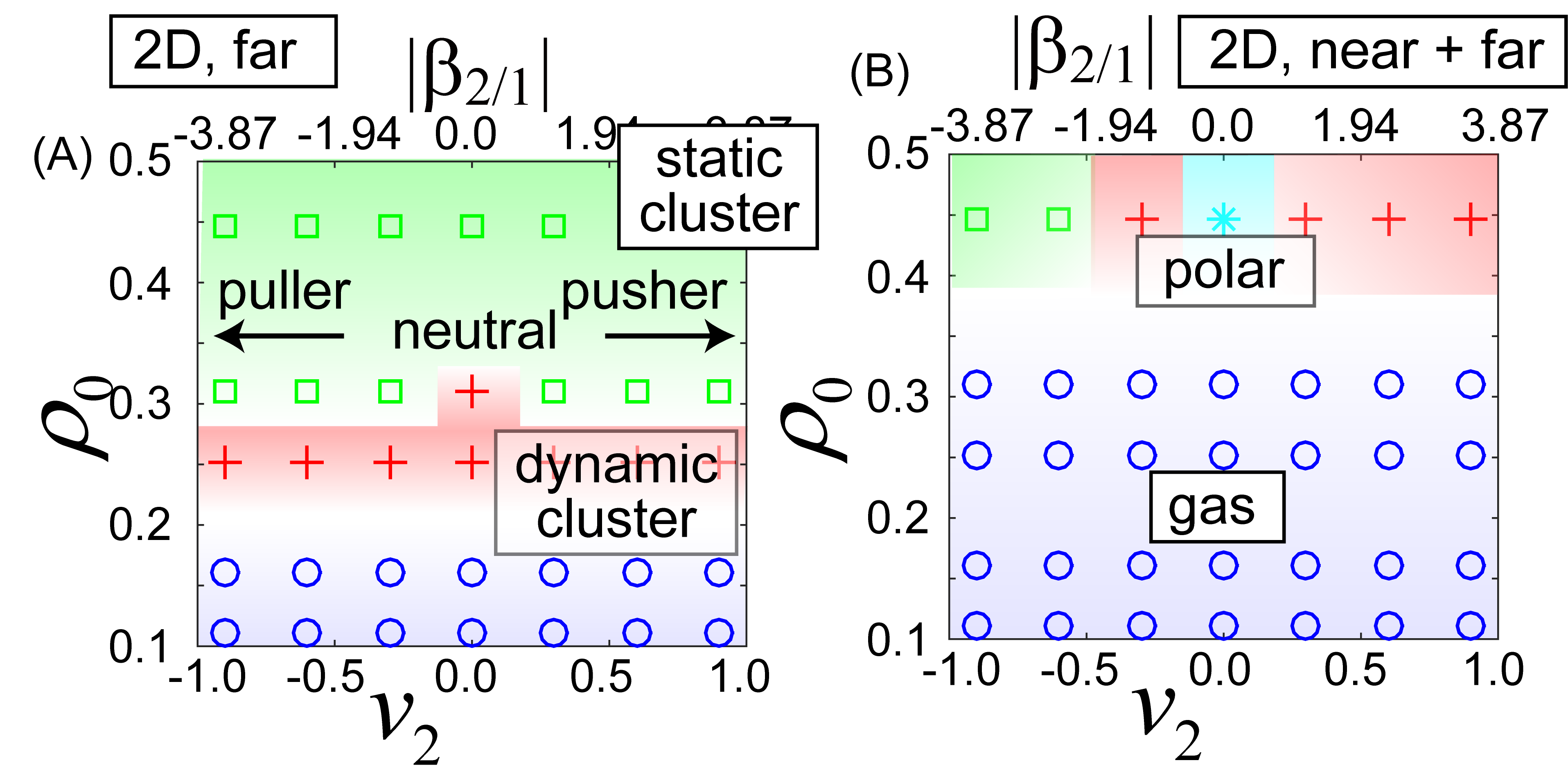}
}
 \caption{
 (Colour Online) The state diagram of squirmers with density ($\rho_0$) and dipolar strength ($v_2$) with (A) far field
 hydrodynamic interactions and (B) both far field and near field in 2D
 systems with $N=512$ particles.
 \label{fig.Phase.Diagram2D}
}
\end{center}
\end{figure}

We also carried out simulations with 2D hydrodynamic interactions using
the method outlined in sect.\ref{sec.2D.squimers}.
  The polar state also appears here  although it occurs at higher densities than the
  system with 3D hydrodynamics as shown in fig.~\ref{fig.Phase.Diagram2D}(B).
 At  intermediate densities, dynamic clusters appear. 
  This is in agreement with~\cite{Leoni:2010a}  which showed a gap between the dimension of the interaction
and the dimension of  the system led to a weaker suppression of alignment.
  Similar to 3D systems, the far-field interaction does not lead to 
polar order even for neutral swimmers as shown in
fig.~\ref{fig.Phase.Diagram2D}(A).
In this case, static clusters appear at higher densities while the
system is disordered at low densities.
The near-field interaction results in  polar order for neutral swimmers.

Swimmers only with the near-field interaction also show  polar order
(fig.~\ref{fig.analysis2}(B)).
In contrast to 3D systems, polar order only appears when the density is
high.
The emergence of polar order is insensitive to system size and polar order is present for the largest system size simulated: 
$N=32768$ particles.
This is analogous to what was found in 3D systems.

 \section{Swimmer collisions and polar order}
\label{sec.collision}
 
 \subsection{Two-body collisions}

\begin{figure}[htbp]
 \begin{center}
    \resizebox{0.5\textwidth}{!}{%
  \includegraphics{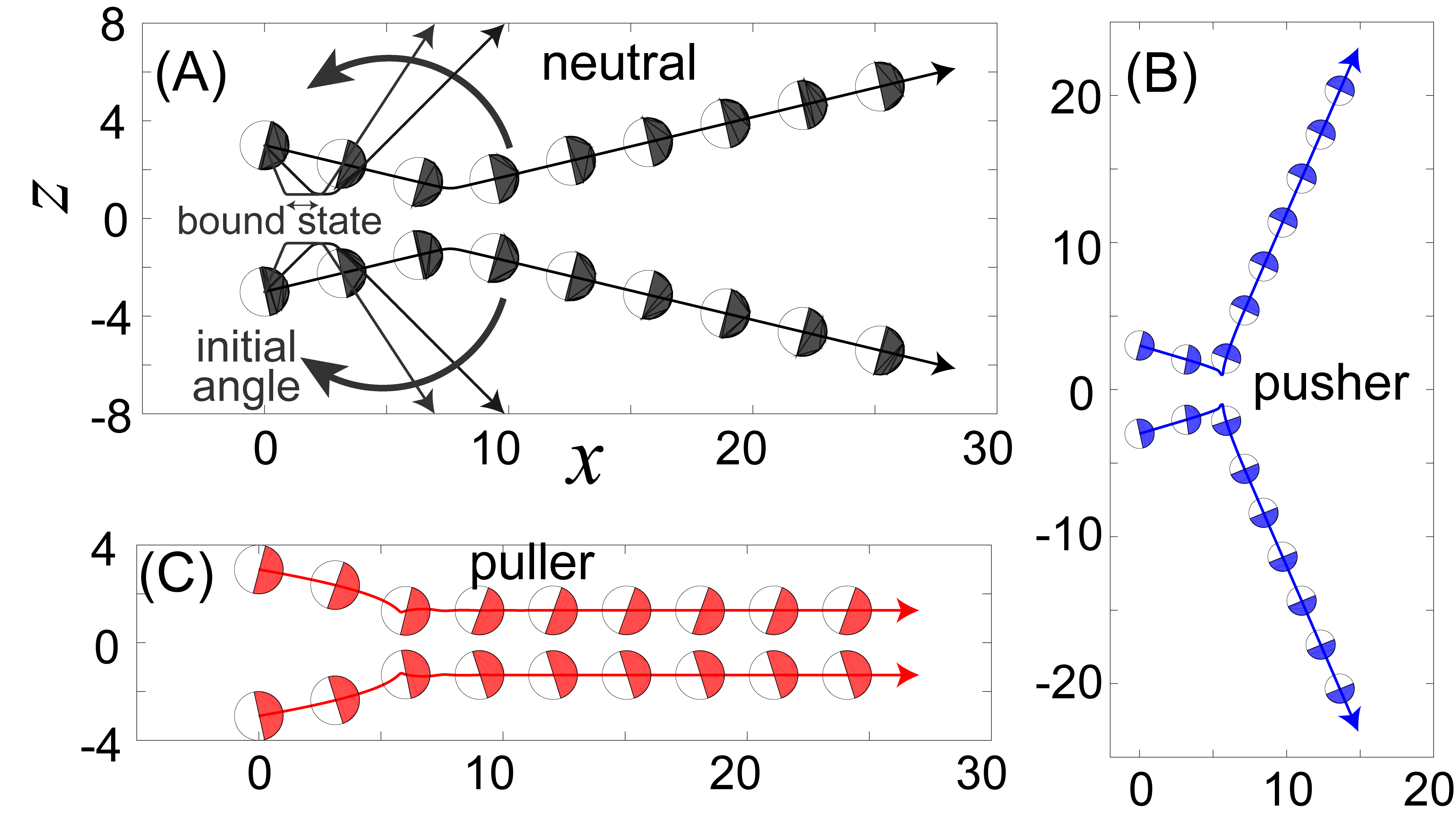}
}
 \caption{
 (Colour Online)
 \NYY{
 Trajectories  for  (A) neutral ($v_2=0$), (B) pusher ($v_2=0.9$), and
 (C) puller ($v_2=-0.9$) swimmers.
 For the neutral swimmers, the trajectories with larger initial angles
 are also shown.
 The directions of the trajectories are shown by arrows.
 }
\label{fig.two.body1}
}
\end{center}
\end{figure}

\tll{
In this section, we investigate the two body interaction in detail to
get insights into the  emergence of collective behaviour leading to polar order.
Figure \ref{fig.two.body1} shows some trajectories of
two `colliding' squirmers, whose initial orientation would lead to them approaching other  in the absence of interactions.
The hydrodynamic (lubrication and far-field) interactions lead to a reorientation of the swimmers which eventually move in separate directions. 
We call this process a `collision'.
If the separation between the initial angles is the same as the separation between the final angles then the {\em collision} is said to be {\em symmetric}.
The neutral swimmer has symmetric collisions when the initial angle is
small while there is a small asymmetry that develops when the initial angle becomes large, leading to smaller angular separations after collisions.
The final angles of the pusher (puller) are larger (smaller) than
the initial angles.
As the initial angle increases, neutral swimmers develop bound states
in which two particles stay a close distance (\NY{fig.~}\ref{fig.two.body1}(A)) from each other for an extended period during which lubrication interactions dominate before eventually separating. These bound states are the source of the alignment and development of polar order observed.
This is because the bound states are robust, i.e. remain stable even after additional collisions with other swimmers while the other post collision states are unstable to future collisions.
}

\begin{figure}[htbp]
 \begin{center}
    \resizebox{0.5\textwidth}{!}{%
  \includegraphics{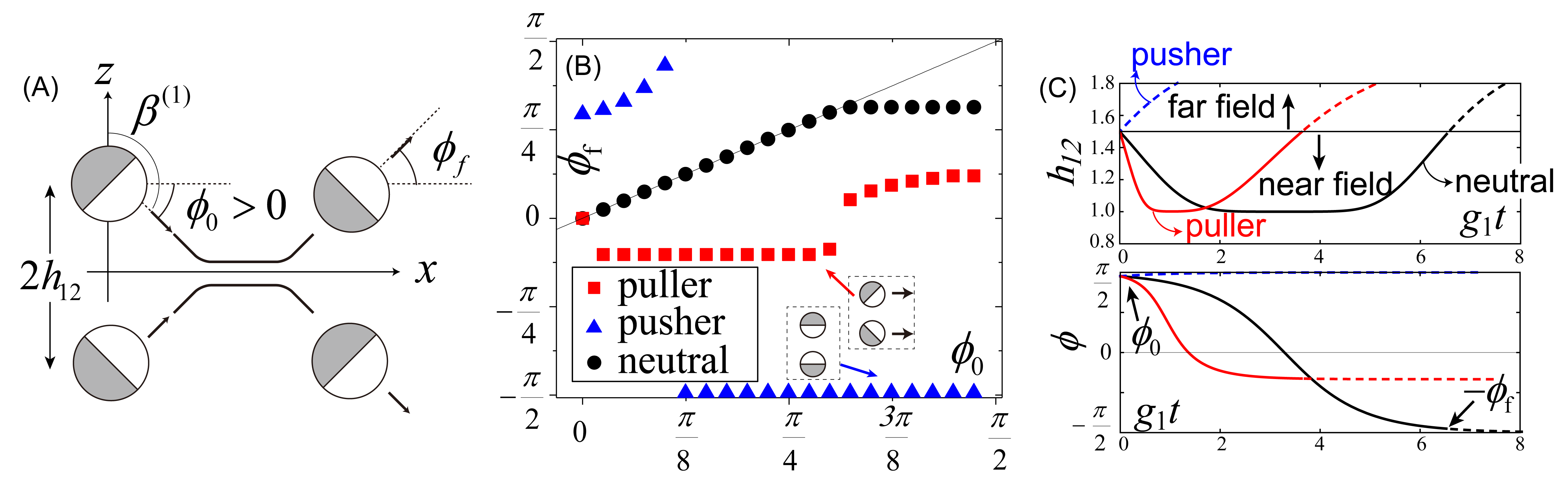}
}
 \caption{
 (Colour Online)
\NYY{
 (A) Schematics of two-body collisions.
 (B) The incidence $\phi_0$ and reflection $\phi_f$ angles
 for symmetric collisions.
 The solid line shows $\phi_0 = \phi_f$.
 The negative $\phi_f$ corresponds to the bound states schematically
 drawn in the insets.
 (C) The dynamics of the separation $h_{12}(t)$ and the angle $\phi(t)$ in the near-field region
 obtained from \NY{eq.~}(\ref{near.omega}) for $g_2/g_1=1$.
 Motion outside the near-field region is indicated by dashed lines.
 }
\label{fig.two.body}
}
\end{center}
\end{figure}

\tll{
In order to quantify the dynamics of collisions, we consider
trajectories with reflection symmetry about the $xy$-plane shown in
\NY{fig.~}\ref{fig.two.body}(A).
 The position and orientation of the two particles are expressed by
$\beta_1 = \pi/2 + \phi$ and $\beta_2 = \pi/2 - \phi$,
$x^{(1)}=x^{(2)}$, and $h_{12} = (z^{(1)} - z^{(2)})/2$ .
}
Due to the near-field interaction, alignment and transient bound
states appear as shown in fig.~\ref{fig.two.body}(B) and (C).
For swimmer separations where near-field effects dominate, the
orientational dynamics is
\begin{align}
 \dot{\phi}
 &= -g_1 \cos \phi + g_2 \sin
 2\phi
 \label{two.body.meanangle}
\end{align}
 with $g_1 \sim u_0 /R$ and $g_2 \sim v_2/R$.
A particle with incident angle $\phi_0$ at the outer boundary of
the near field region at $t=0$ rotates so that
$\phi(\tau)=0$ at the collision time $t=\tau$.
For the neutral swimmer $g_2=0$, we have $g_1
\tau = \log (1-\tan \phi)/(1+\tan \phi)$.
\tll{
The trajectory is shown in fig.~\ref{fig.two.body}(C)).
When the initial angle is large enough, a neutral swimmer spends a
long time bound to another.
}
%
%
%

For the far-field-only system, two  pushers  
($v_2>0$) align while two pullers ($v_2<0$) stick to each other.
However, these states are unstable to rotational fluctuations
induced by the fact that a swimmer typically experiences collisions with many particles along its trajectory. 
Hence the system does not develop any polar order.
%
The near-field interaction leads to alignment and transient bound
states as shown in fig.~\ref{fig.two.body}(B) and (C).
Neutral swimmers ($v_2=0$), have the longest residence times 
and the weakest rotational fluctuations due to the
absence of rotations from far field interactions.
For small incident angles ($\phi_0$), the reflection angle ($\phi_f$)  increases linearly with incidence angle, 
$\phi_f \simeq \phi_0$, leading to symmetric collisions.
However as the incidence angle becomes larger, the reflection angle no longer increases with the incident angle 
and we observe a {\em saturation} of the reflection angle at a value $|\phi_s| \lesssim \pi/4$ as shown in fig.~\ref{fig.two.body}(C). 
This asymmetry between incident and reflection angles 
($\mean{\phi_f} <  \mean{\phi_0}$, see fig.~\ref{fig.two.body}(C)) is what leads eventually to alignment.
Similar behaviour is seen for  pushers and pullers,
 however, shorter residence times and stronger rotational fluctuations
 destroy the polar state for both of them. 

In simulations, the details of this process depend weakly on the contact
interaction; the choice of the interaction potential leads only to slight shifts of
saturation angles.
Nevertheless, our conclusions about the collective behaviours and phase boundaries are independent of the choice of potential.
We have used the truncated Lennard-Jones repulsive interaction as well as other types of interaction, e.g. \NY{eq.~}(\ref{potential.Brady}).
They all result in qualitatively the same behaviour with
a slightly shifted saturation angle and boundaries between different types of collective behaviour.
We note that the saturation angle ($|\phi_s| \approx \pi/4$) in fig.~\ref{fig.two.body}(E) originates
from direct contacts between squirmers via repulsive forces from the interaction potential.
Without the repulsive interaction, \NY{eq.~}(\ref{two.body.meanangle}) implies
that  incident and  final angles are the same as shown in
fig.~\ref{fig.two.body}(F).
This suggests that the polar order we observe for  neutral and near neutral swimmers is induced by
saturation of a final angle with respect to an incident angle, which
arises from the
loss of its memory of an incident angle due to the contact interaction.

\subsection{Many-body effects}

Although our model includes only pairwise interactions, combining them with each other, 
results in many-body effects which become relevant for a non-dilute
suspension.
In fact, for a dense suspension where $\xi - 2  \lesssim 1 $, the interaction
is dominated by the lubrication interaction between two swimmers which is well 
approximated by a sum of two-body interactions.

\begin{figure}[htbp]
 \begin{center}
    \resizebox{0.5\textwidth}{!}{%
  \includegraphics{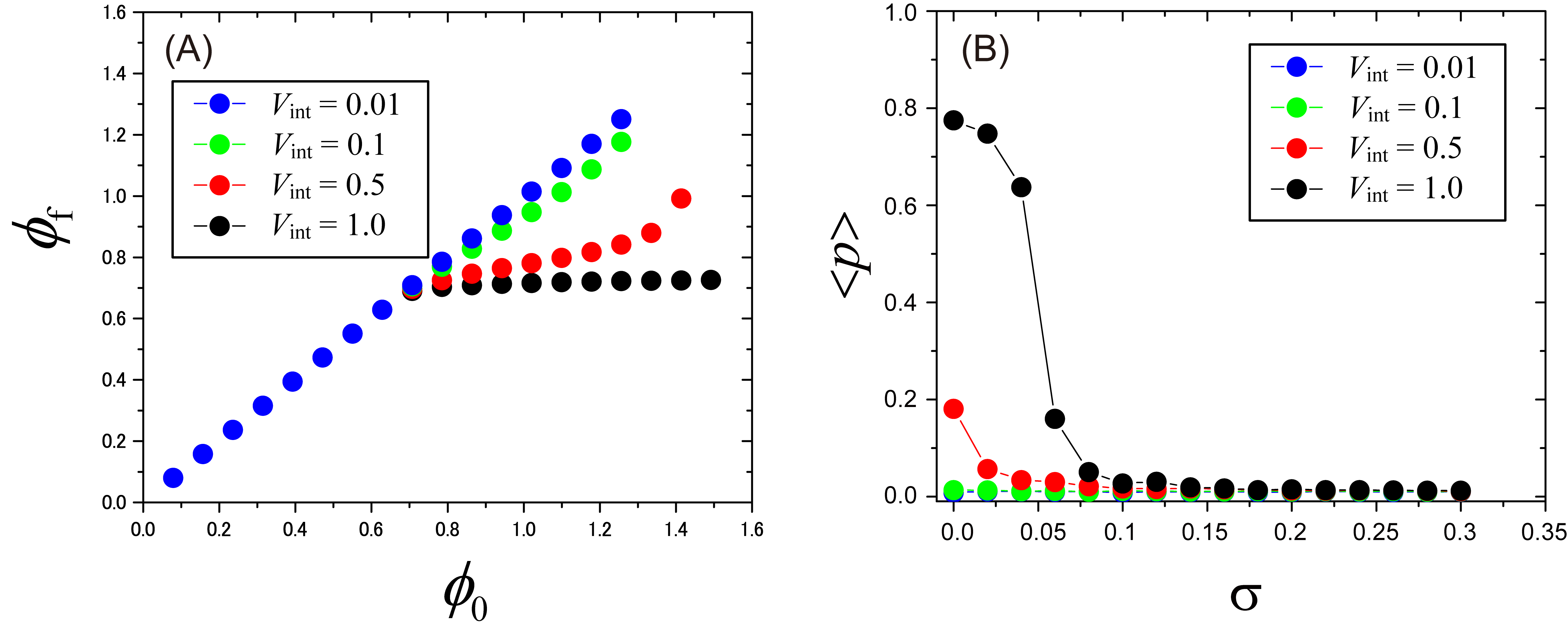}
}
 \caption{
 (Colour Online) (A) Incident and the final angles of two-body collisions
 between two swimmers described by the model with the interaction \NY{eq.~}(\ref{ang.vel.vicsek}). (B) The mean polar order $\mean{p}$ as a function of the
 strength of noise $\sigma$ at $N=8192$ and $\rho_0=0.16$
 ($\xi = 4.43$) corresponding to (A).
\label{fig.vicsekstyle}
}
\end{center}
\end{figure}

To understand how these many-body effects give rise to collective behaviour, we have carried out
 numerical simulations of a Vicsek-style model, in which the interactions in
the angular direction (fig.~\ref{fig.vicsekstyle}(A)).
The interaction is given by
  \begin{align}
    \omega^{(i)}
 &=
 \frac{1}{2}
 \sum_{j \neq i}
 \left[
 \sin \left(
 \beta^{(i)}
 - \theta_{ji}
 \right)
 +
 \sin \left(
 \beta^{(j)}
 - \theta_{ji}
 \right)
   \right]
   \label{ang.vel.vicsek}
  \end{align}
 We are able to reproduce the same polar-disorder phase transition by increasing
 the amplitude of noise as in fig.~\ref{fig.vicsekstyle}(B).
 Therefore, the detailed form of the hydrodynamic interactions in the translational direction
 given by \NY{eq.~}(\ref{trans.vel}) are not essential for the  development of polar order, and
 only the effect of the hydrodynamic interactions on the orientational dynamics in \NY{eq.~}(\ref{ang.vel}) are necessary.
One critical difference with the conventional Vicsek model is that here, the
excluded volume interactions are necessary to generate polar order; without them 
  the system remains the gas state even at  lower values of the  noise amplitude  and with
  strong rotational interactions.
  \tll{
  Polar order appears irrespective of the choice of repulsive
  interactions.
  }
  Here in order to see the effect of excluded volume interaction, we use
 the repulsive interaction as ${\bf F}_{ij} = V_{\rm
 int} (r_{ij} - 2R) \hat{\bf r}_{ij}$.

\subsection{Hydrodynamic argument}

The existence of polar order is fundamentally surprising because of the apparent contradiction with the well accepted generic instability of polar/nematic order of wet active matter~\cite{Simha:2002,Marchetti:2013}. An explanation of how this is possible can be found using a simple argument based on a generalized hydrodynamic description of the conserved quantities and spontaneously broken continuous symmetries of the system~\cite{Marchetti:2013}.  It is noteworthy that it emphasizes the importance of collisions of the swimmers with each other as essential for the formation of polar order. 

We begin by constructing a two-fluid model for the system, for the suspending fluid (volume fraction, $1-\phi$) and the active particle (squirmer) 'network' (volume fraction $\phi$). 
We define a local displacement variable $\bfu (\bfr,t)$ for the active particle network which is related  to the density variations of squirmers.
The Newtonian fluid it is suspended in is characterised by a velocity field $\bfv (\bfr,t)$. Finally we identify a  {\em local} polar order parameter,  $\bfp (\bfr,t)$. An isolated squirmer swims with velocity $u_0 \bfp$ relative to the background fluid.
The fluid obeys equation,
\beq
\rho_f \dot \bfv = \eta \nabla^2 \bfv - \nabla P + \nabla \cdot {\bsigma}^a - \bff^c
 \label{eq:2fl1}
\eeq
where $\rho_f = (1-\phi) \rho$ is the average density of fluid, $\bff^c \propto \rho_a^2$ is the force on a squirmer due to collisions with others, where $\rho_a = \phi \rho$ is the average density of active particles. %
The fluid is taken as incompressible,
 \bem
 \nabla \cdot \bfv = 0 \; . \; 
 \eem
The  active stress $\bsigma^a$ is given by 
 \bem
 \sigma^a_{ij} = {\nu p_i p_j} \; ,
 \eem
 where $\nu \propto v_2$.

 The active 'network' has an equation of motion which indicates that the velocity, $\dot \bfu (\bfr,t)$ of the active particles at $\bfr$ is given by 
 \beq
 \dot \bfu (\bfr,t) =  \bfv (\bfr,t) + u_0 \bfp (\bfr,t) - \bfF /\zeta
 \label{eq:2fl2}
 \eeq
 where $\bfF(\bfr,t)$ is the total force density on the active particles at point $\bfr$, and $\zeta$ is the  slope of the force-velocity curve of the squirmer, i.e. in the absence of external forces (here due to collisions), the squirmers move with speed $u_0$.

In addition there is the equation of motion for the polar director 
\beq
\dot \bfp = - u_0 \bfp \cdot \nabla \bfp +  \bomega \cdot  \bfp + \gamma \bfe \cdot \bfp + K \nabla^2 \bfp + \cdots
\eeq
where $\omega_{ij} = \frac12 ( \partial_i v_j - \partial_j v_i)$,
$e_{ij} = \frac12 ( \partial_i v_j + \partial_j v_i)$ and \tlll{$K$ is an effective
Frank elastic constant that assigns a cost to distortions of the local polar order and in general will include non-equilibrium contributions~\cite{Marchetti:2013,TjiptoMargo:1992}.
}
 
It is illustrative to consider the system at low density as $\rho_a \;  \rar \; 0$. Then we find $\bff^c=0$ and performing a linearised analysis about the 
homogeneous state, $\dot\bfu = \dot\bfu_0, \bfp=\bfp_0$ , $ \bfv_0 = \dot\bfu_0 - u_0 \bfp_0 $
we obtain the generic instability of the polar state~\cite{Simha:2002}.

Now let us switch on $\bff^c \ne 0$.
 The force $\bff^c$ leads to slowing down of the swimmer, taking \NY{eq.~}(\ref{eq:2fl2}) above when gradients of $\bfu$ vanish :
 \beq
 \bff^c = \zeta \left( \dot \bfu - \bfv - u_0 \bfp \right) \quad \Rar \quad \dot \bfu = \bfv + u_0 \bfp + \bff^c / \zeta \; , 
 \eeq
 i.e. when $\bff^c=0$, the squirmer swims at its free speed.

We can replace $\bff^c$ by $\zeta \left( \dot \bfu - \bfv - u_0 \bfp \right)$ in \NY{eq.~}(\ref{eq:2fl1}).
So now performing the linearised expansion $\bfp = \bfp_0 + \delta
\bfp(\bfr,t), \bfv = \bfv_0 + \delta \bfv (\bfr,t), \bomega = \bomega_0
+ \delta \bomega(\bfr,t), \bfe = \bfe_0 + \delta \bfe (\bfr,t)$ around $ \bfv_0 = \dot\bfu_0 - u_0 \bfp_0 $,
\beqa
0 &=& \eta \nabla^2 \delta \bfv + \zeta \delta \bfv - \nabla P + \nabla \cdot {\bsigma}^a \; ; \; \nabla \cdot  \delta \bfv = 0 \;,  \\
 \partial_t \delta \bfp &=& - u_0 \bfp_0 \cdot \nabla \delta \bfp +
 \delta \bomega \cdot  \bfp_0 + \gamma \delta \bfe \cdot \bfp_0 + K \nabla^2 \delta \bfp \ldots \; , 
 \eeqa
with a finite screening length $\xi \sim \sqrt{\eta/\zeta}$ which weakens the generic instability from a long-wavelength to a finite wavelength instability and stabilizes the polar state on long lengthscales. Hence a comparison between the screening length $\xi$ and the active lengthscale, $\sqrt{K/|\nu|}$~\cite{Voituriez:2005}, allows us to determine the onset of polar order for $|\nu| < \nu_c = K / \xi^2$, i.e. for swimmers that are close to neutral.
\tlll{In this long-wavelength limit, the appearance of 
polar order is symmetric  about $\nu=0$.
However, short range hydrodynamics corresponding to  higher-order terms in spatial gradients can easily break that symmetry.
}

\section{Discussion and summary}
\label{sec.summary}

\tlll{In summary, we have re-analyzed the collective behaviour of self-propelled particles (squirmers) taking account of 
the effects of hydrodynamics when the swimmers are close to one another.
We find the existence of a polar ordered phase whose stability is dominated by short-range interactions
arising from lubrication forces between two squirmers. The global phase
separation of active particles with repulsive interactions, 
is suppressed by hydrodynamic interactions and we observe instead
dynamic clusters of finite size for a large range of intermediate
densities.
We see gel-like extended state at high enough densities.}

We emphasize that our goal is to develop an {\em approximate} method of dealing with hydrodynamics 
of squirmers, that can efficiently simulate large enough numbers of swimmers  
to capture accurately the macroscopic behaviour of wet active matter. 
Polar order has been observed in earlier 
simulations with hydrodynamics, but due to the computationally expensive nature of fluid mechanical calculations, these were limited by small numbers of particles and it remained an open question if the observed phenomena were simply finite size effects. 
Due to limitations in computational power, there is a trade-off that must be made between how accurately the hydrodynamic flow field is resolved and increasing the number of particles. For Navier-Stokes solvers like Lattice Boltzmann or Multi-Particle Collision Dynamics, this is typically done by 
limiting the resolution of the flow field to a scale comparable to the squirmer size so that the behaviour of particles in very close proximity is not 
treated accurately; such models are best for particles at intermediate separations. Here we develop an alternative approach with different strengths; developing a description that is very accurate when the particles are very close and very far apart but which performs less well for intermediate separations.
A further strength of our method is the ability to switch of different contributions to the motion to identify the dominant mechanisms behind each macroscopic phenomena. With this model we are able to simulate large numbers of particles and confirm that the qualitative behaviour seen by small simulations remains even for much larger systems.  

Nevertheless, we do find some quantitative differences with the previous numerical results.
Of particular interest is the transition induced by puller slip flow (negative
$v_2$), namely it has been reported that the amount of polar order
for pullers near the neutral swimmer ($v_2=0$) is larger than
that of pushers which our results do not show.
\tlll{Interestingly, in our simulations, see 
 fig.~\ref{fig.v2}(A,C), we find an asymmetry in the appearance of polar order about $v_2=0$;
that is, the amount of polar order observed for pushers ($v_2>0$) is different to 
 pullers ($v_2<0$).
For pushers, peak polarity is slightly shifted from $v_2=0$, and the amount of polar order decays
rapidly to zero beyond a critical force dipole magnitude.
For pullers, the polar order first decays rapidly to $\mean{p}\simeq
0.2$, then decays slowly to zero as $v_2$ increases, particularly
for smaller systems.
This slow decay does not appear in the near-field-only
systems.
This indicates the importance of the far-field hydrodynamics, and
also implies that the polar order for pullers at intermediate values 
of $v_2$ is influenced by medium-scale hydrodynamic interactions, which are only crudely approximated within our approach. 
}
It has also been reported that a density wave appears for pullers
\cite{Alarcon:2013}.
We have seen this in the purely two-dimensional system (2D with 2D interactions) and
quasi-two-dimensional system (2D with 3D interactions) with noise, but not for pullers in the
quasi-two-dimensional system (2D with 3D interactions).
\tlll{We emphasize however that, since each  of the methods used so far has disadvantages, either small number of
particles, inaccurate treatment of the near-field interactions, or intermediate range interactions; that 
further work is needed to clarify some of these issues.} We may speculate however that the behaviour above seen using the 
Lattice Boltzmann Method to simulate pullers
is dominated by intermediate-ranged, many-body hydrodynamic interactions,
which are not well captured by our method.

Since our treatment of hydrodynamics is approximate, particularly for
intermediate distances between swimmers, a detailed comparison with high resolution full hydrodynamic simulations is 
required in the future. In this direction, it would  be useful to investigate the effect of
 lubrication forces on the simulation techniques accurate for intermediate to
 long separation between squirmers, such as the Lattice Boltzmann Method
 and Multi-Particle Collision Dynamics.
On the other hand, it would also be interesting to include intermediate-distance
 interactions more accurately in our model by using particle-mesh-type
 simulations, which has been used in particle simulations with electrostatic interactions~\cite{Maggs:2002}.

\section*{Acknowledgments}
The authors are  grateful to S. Fielding, T. Ishikawa and R. Golestanian for
 helpful discussions.
  NY acknowledges the support by JSPS KAKENHI Grant Numbers JP16H00793
 and 17K05605.
 TBL is supported by BrisSynBio, a BBSRC/EPSRC Advanced Synthetic Biology
 Research Center (grant number BB/L01386X/1).
We would like to thank the Isaac Newton Institute for Mathematical Sciences, Cambridge, for support and hospitality during the programmes, ``The Mathematics of Liquid Crystals'' and ``Dynamics of active suspensions, gels, cells and tissues'' where work on this article was started.

\NY{
\section*{Author contribution statement}
NY and TBL developed the theoretical model. NY performed the
simulations. Both authors interpreted the results and wrote the paper.
}

\appendix

\section{Coefficients of vector spherical harmonics}
\label{sec.appendix.SH}

Since the expansion of the slip velocity in terms of the vector spherical
harmonics is carried out with the $z$-axis in a (fixed) laboratory frame,
the coefficient of each spherical harmonic in the slip velocity $v_{lm}$ varies with rotation of the particle.
There is
a one-to-one map from the angles to the coefficients $\{ \alpha, \beta \}
\mapsto v_{lm}$.
When $v_{lm}$  has only the first, $l=1$ mode , we may define a
vector ${\bf p}$ with components $v_{1,-1}$, $v_{1,0}$, and $v_{1,1}$.
The second mode, $l=2$ can be represented by a second-rank tensor, with 
higher modes correspond to higher-rank tensors \cite{Yoshinaga:2014}.
In terms of spherical harmonics, these are given by $(2l+1)$-dimensional vectors.
Here $v_{lm} = (v_{l,-l}, \cdots v_{l,0}, \cdots, v_{l,l})$ is expressed by the polar angle $
 {\bf p}^{(i)} 
=
\left(
\cos \alpha \sin \beta,
\sin \alpha \sin \beta,
\cos \beta
\right)
$.
  as
\begin{align}
v_{1,m}^{(i)} 
&=
 v_1^{(i)}
 \begin{pmatrix}
\frac{1}{\sqrt{2}} \sin \beta^{(i)} e^{i \alpha^{(i)}}
\\
  \cos \beta^{(i)}
  \\
-\frac{1}{\sqrt{2}} \sin \beta^{(i)} e^{-i \alpha^{(i)}}
 \end{pmatrix}
\end{align}
and
\begin{align}
v_{2,m}^{(i)} 
 &=
v_2^{(i)}
\begin{pmatrix}
 \frac{\sqrt{6}}{4} \sin^2 \beta^{(i)} e^{2 i \alpha^{(i)}}
 \\
 \sqrt{\frac{3}{2}} \sin \beta \cos \beta^{(i)}  e^{i \alpha^{(i)}}
 \\
 \frac{3}{2} \cos^2 \beta^{(i)} - \frac{1}{2}
 \\
- \sqrt{\frac{3}{2}} \sin \beta^{(i)} \cos \beta^{(i)} e^{-i
 \alpha^{(i)}}
 \\
 \frac{\sqrt{6}}{4} \sin^2 \beta^{(i)} e^{-2 i \alpha^{(i)}}
 \label{v2.5d.vector}
\end{pmatrix}
.
\end{align}

   \section{Cartesian tensors}
\label{sec.Cartesian}
While we have written out our flow fields in terms of vector spherical harmonics, they  may equivalently be expressed in terms of 
  Cartesian tensors. 
  To do this, we define the  {\it intrinsic} velocity ${\bf U}_0$, the {\it intrinsic} angular
velocity, and the {\it intrinsic} strain flow arising from slip velocity, \NY{eq.~}(\ref{janus.intro.slip}) by
\begin{align}
 {\bf U}_0
 &=
 - \frac{1}{4 \pi R^2}
 \int
 {\bf v}_s
 dS
 \\
 {\bf \Omega}_0
 &=
 -\frac{3}{8 \pi R^3}
 \int
{\bf n} \times {\bf v}_s
 dS
 \\
 {\bf E}_0
 &=
 -\frac{3}{8 \pi R}
 \int
 \left[
 {\bf n} {\bf v}_s
 +  {\bf v}_s  {\bf n}
 \right]
 dS
\end{align}
where ${\bf n}(\theta,\varphi)$ is a normal vector pointing outward a
spherical surface.
In terms of the coefficients in \NY{eq.~}(\ref{janus.intro.slip}), these quantities are expressed as
\begin{align}
 {\bf U}_0
 &=
 - \frac{1}{4 \pi }
 4 \sqrt{\frac{\pi}{3}}
 \begin{pmatrix}
 \frac{v_{1,-1} - v_{1,1}}{\sqrt{2}} \\
 -i \frac{v_{1,-1} + v_{1,1}}{\sqrt{2}} \\
 v_{1,0}
 \end{pmatrix}
\end{align}
\begin{align}
 {\bf \Omega}_0
 &=
 \frac{3}{8 \pi R}
  4 \sqrt{\frac{\pi}{3}}
 \begin{pmatrix}
 \frac{w_{1,-1} - w_{1,1}}{\sqrt{2}} \\
 -i \frac{w_{1,-1} + w_{1,1}}{\sqrt{2}} \\
 w_{1,0}
 \end{pmatrix}
\end{align}
\begin{align}
  \frac{ {\bf E}_0 }{ -\frac{3R}{8 \pi}
  4 \sqrt{\frac{3\pi}{10}}
}
 =
 \begin{pmatrix}
   v_{2,\pm 2} - \sqrt{\frac{2}{3}} v_{2,0}&
  i  \tilde{v}_{2,\pm 2} &
    -  \tilde{v}_{2,\pm 1}
  \\
    i  \tilde{v}_{2,\pm 2} &
  -  v_{2,\pm 2} - \sqrt{\frac{2}{3}} v_{2,0} &
  -i  v_{2,\pm 1} &
  \\
  -   \tilde{v}_{2,\pm 1} &
    -i v_{2,\pm 1} &
 2 \sqrt{\frac{2}{3}} v_{2,0}
 \end{pmatrix}
\end{align}
where
\begin{align}
 v_{2,\pm 2}
 &=
 v_{2,2} + v_{2,-2}
 \\
  \tilde{v}_{2,\pm 2}
 &=
 v_{2,2} - v_{2,-2}
 \\
 v_{2,\pm 1}
 &=
 v_{2,1} + v_{2,-1}
 \\
 \tilde{v}_{2,\pm 1}
 &=
 v_{2,1} - v_{2,-1}
\end{align}
Using the orientation of a swimmer, ${\bf p}$, in \NY{eq.~}(\ref{swimmer.axis}), the 
{\it intrinsic} velocities are expressed as
\begin{align}
 {\bf U}_0
 &=
 - \frac{v_1}{\sqrt{3\pi}} {\bf p}
 \\
 {\bf \Omega}_0
 &=
- \frac{w_1}{2}
 \sqrt{\frac{3}{\pi}}
 {\bf p}
 \\
 {\bf E}_0
 &=
 - \frac{3 R v_2}{4}
 \sqrt{\frac{3}{\pi}}
 \left(
{\bf p}{\bf p} - \frac{1}{3} {\bf I}
 \right)
\end{align}
The physical meaning of these quantities is clear from Faxen's laws.
The force, torque, and stress for a {\it fixed} isolated swimmer is expressed as
\begin{align}
 {\bf F}^{(a)}
 &=
 - 6 \pi \eta R {\bf U}_0
 \\
 {\bf T}^{(a)}
 &=
 -8 \pi \eta R^3
 {\bf \Omega}_0
 \\
 {\bf S}^{(a)}
 &=
- \frac{20}{3} \pi \eta R^3
 {\bf E}_0
 .
\end{align}
These are active force, torque, and stress discussed in
sect.~\ref{sec.pariwise}.
Note that the active stress, $ {\bf S}^{(a)}$ does not lead to motion
and rotation of an isolated swimmer.

We will rewrite the velocity field in Cartesian coordinates and the
{\it intrinsic} flow, ${\bf U}_0$, ${\bf \Omega}_0$, and ${\bf E}_0$
described above.
We expand the velocity field in terms of the modes $l$ as
\begin{align}
 {\bf v} ({\bf r})
 &=
 \sum_{l=1}^{\infty}
 {\bf v}_l ({\bf r})
 .
\end{align}
The first mode is 
\begin{align}
 {\bf v}_1 ({\bf r})
 &=
 \left(
\frac{R}{r}
 \right)^3
 \left[
{\bf U}_0 \cdot {\bf n}{\bf n}
 - \frac{1}{2}
 \left(
{\bf U}_0 \cdot {\bf t} {\bf t}
 +
{\bf U}_0 \cdot {\bf b} {\bf b} 
 \right)
 \right]
 \nonumber \\
 &=
 \left(
\frac{R}{r}
 \right)^3
 \left[
 \frac{3}{2}
{\bf U}_0 \cdot {\bf n}{\bf n}
 - \frac{1}{2}
{\bf U}_0
 \right] 
 .
\end{align}
The second mode is given by
\begin{align}
 {\bf v}_2 ({\bf r})
 =&
 3 \left[
 \left( \frac{R}{r} \right)^2
 - \left( \frac{R}{r} \right)^4
 \right]
 {\bf E}_0 \odot_2 {\bf n} {\bf n} {\bf n}
 \nonumber \\
& + \left( \frac{R}{r} \right)^4
 \left(
 {\bf E}_0 \odot_2 {\bf t} {\bf t} {\bf t}
 +
  {\bf E}_0 \odot_2 {\bf b} {\bf b} {\bf b}
 \right)
\end{align}
where we denote contraction of two indices by
\begin{align}
 {\bf A} \odot_2 {\bf B}
 &=
 A_{ij} B_{ij}
 .
\end{align}

\section{Axisymmetric motion in 3D}
\label{sec.app.axisymmetric.3D}

First we consider the motion in $z$-direction.
In polar coordinates, the velocity field is described using the
stream function as
\begin{align}
{\bf v} 
&=
\left(
\frac{1}{r} \pdiff{\psi}{z},
0,
- \frac{1}{r} \pdiff{\psi}{r}
\right)
.
\end{align}
The stream function satisfies the biharmonic-like equation
\begin{align}
E^2 E^2 \psi 
 &= 0
 \label{SM.near.sym.biharmonic}
\end{align}
with the operator $E^2 = \pdiff{^2}{r^2} - \frac{1}{r} \pdiff{}{r}
+ \pdiff{^2}{z^2}$.
We expand the stream function in  $\epsilon$ as
\begin{align}
 \psi
&=
 \epsilon \psi_0 + \epsilon^2 \psi_1 + \cdots
 \label{SM.near.3D.sym.stream}
\end{align}
with the operator also expanded as
\begin{align}
E^2 
&=
\frac{1}{\epsilon^2} \pdiff{^2}{\tilde{Z}^2}
+ \cdots
.
\end{align}
For the near field, the velocity fields are at the lowest order
$
v_r 
\sim
\mathcal{O}(\epsilon^{-1/2})$ and 
$
v_z
 \sim
\mathcal{O}(\epsilon^0)
$.

The passive force acting on the sphere given a translational velocity $u_z$ is \cite{Stimson:1926,Cooley:1969}
\begin{align}
\frac{ F_{z1}^{(p)} }{6 \pi \eta R}
&\simeq
- \frac{1}{4 \epsilon} 
\left(
u_z^{(i)} - u_{z}^{(j)}
\right)
-  
\frac{1}{2}
\left(
u_z^{(i)} + u_{z}^{(j)}
\right)
\mathcal{I}_{z}^{(p)}
\\
\frac{ F_{z2}^{(p)} }{6 \pi \eta R}
& \simeq
 \frac{1}{4 \epsilon} 
\left(
u_z^{(i)} - u_{z}^{(j)}
\right)
- 
\frac{1}{2}
\left(
u_z^{(i)} + u_{z}^{(j)}
\right)
\mathcal{I}_{z}^{(p)}
\end{align}
Here we have two terms; one is a singular term obtained from
asymptotic analysis of the problem (b), and the other term is $\mathcal{O}(\epsilon^0)$,
which corresponds to the problem (f) in fig.~\ref{fig.sec1.8problems}.
Extending the matched asymptotic expansion, the calculation of the
$\mathcal{O}(\epsilon^0)$ term may be possible. However,
instead, the patching procedure is often used; $\mathcal{O}(\epsilon^0)$
term is replaced by the value when two sphere is touching each other \cite{Goldman:1967a}.
This value could be obtained either from numerical results or from the
analytical calculation using tangent-sphere coordinates with some
truncation.
In fact, the unknown constant $\mathcal{I}_{z}^{(p)}$ is computed as
$
 \mathcal{I}_{z}^{(p)}
=
\mathcal{O}(\epsilon^0)
\simeq
0.645141
$.
It may also be possible to replace the force acting on the two particles
moving together approximately by that of the isolated particle.
Then we set $\mathcal{I}_{z}^{(p)} = 1$.

For the active force, we expand the stream function instead as \NY{eq.~}(\ref{SM.near.3D.sym.stream})
\begin{align}
 \psi
&=
 \epsilon^2 \psi_0 + \epsilon^3 \psi_1 + \cdots
\end{align}
which is appropriate for the boundary conditions in the
active problem, $v_r \sim \mathcal{O}(\sqrt{\epsilon})$ and $v_z \sim \mathcal{O}(\epsilon)$. 
The force acting on the sphere is computed as
\begin{align}
\frac{F_{z1}^{(a)}}{6 \pi \eta R}
&\simeq
 - \NY{\frac{1}{2}}
  \log \epsilon
 \sum_l
 \NY{\mathcal{N}_{l,0}}
V_{l,0}
+ 
\frac{1}{3}
\mathcal{I}_z^{(a)}
 \mathcal{N}_{1,0}
 W_{1,0}
 \label{axisym.Fz1}
\\
\frac{ F_{z2}^{(a)} }{6 \pi \eta R}
 & \simeq
 \NY{\frac{1}{2}}
  \log \epsilon
 \sum_{l}
 \NY{ \mathcal{N}_{l,0} }
 V_{l,0}
+
\frac{1}{3}
\mathcal{I}_z^{(a)}
 \mathcal{N}_{1,0}
 W_{1,0}
\end{align}
where as in the main text, we have used
\begin{align}
 V_{lm}
 &=
 \frac{l(l+1)}{2}
 \left(
 (-1)^l v_{lm}^{(i)}
 +
 v_{lm}^{(j)}
 \right)
 \\
 W_{lm}
 &=
 \frac{l(l+1)}{2}
 \left(
 (-1)^l v_{lm}^{(i)}
 -
 v_{lm}^{(j)}
 \right) \quad .
\end{align}
We simply replace this force by the force required to fix an
{\it isolated} squirmer by adding an unknown constant
$\mathcal{I}^{(a)}_z \sim \mathcal{O}(\epsilon^0)$.

\section{Non-axisymmetric motion and rotation in 3D}
\label{sec.app.nonaxisymmetric.3D}

When the spheres move with translational velocity in $x$-direction and
with angular velocity in $y$-direction, the velocity field is expressed
in polar coordinates as
\begin{align}
{\bf v} 
&=
\left(
U \cos \varphi,
V \sin \varphi,
W \cos \varphi
\right)
.
\end{align}
The boundary conditions are rewritten as
\begin{align}
U (\tilde{Z}=H_{i}) 
&=
U_{H_{i}}
\nonumber \\
V (\tilde{Z}=H_{i}) 
&=
 V_{H_{i}}
 \label{SM.3D.near.asym.BC}
\\
W (\tilde{Z}=H_{i}) 
&=
 W_{H_{i}}
 \nonumber
 ,
\end{align}
\NYY{
where the subscripts $i=1,2$ correspond to Sphere (i) and (j),
respectively, in fig.~\ref{fig.sm.coordinate}.
}
The velocity on the surface is not necessarily constant as for rigid translational
and rotational motion, but is dependent on $\tilde{\rho}$.
The expansion discussed in the next section assumes the boundary
conditions satisfy
$
 U_{H_1} 
\sim \mathcal{O}(\epsilon^0)
,
\; \;
 U_{H_2} 
\sim \mathcal{O}(\epsilon^0)
,
\; \;
 V_{H_1} 
\sim \mathcal{O}(\epsilon^0)
,
\; \;
V_{H_2}
\sim \mathcal{O}(\epsilon^0)
,
\; \;
 W_{H_1} 
\sim \mathcal{O}(\epsilon^{1/2})
,
\; \;
W_{H_2}
\sim \mathcal{O}(\epsilon^{1/2})
$
.
Following \cite{COOLEY:1968}, the velocity fields are expanded in 
$\epsilon$ as
\begin{align}
P (r,z) 
&=
\epsilon^{-3/2} P_0 (\tilde{\rho}, \tilde{Z}) 
 + \mathcal{O}(\epsilon^{-1/2})
 \\
U(r,z)
&=
U_0  (\tilde{\rho}, \tilde{Z}) 
+ \mathcal{O}(\epsilon)
\\
V(r,z)
&=
V_0  (\tilde{\rho}, \tilde{Z}) 
+ \mathcal{O}(\epsilon)
\\
W (r,z)
&=
\epsilon^{1/2}
W_0  (\tilde{\rho}, \tilde{Z}) 
+ \mathcal{O}(\epsilon^{3/2})
.  
\end{align}
At  lowest order in $\epsilon$, the Stokes equation becomes \cite{COOLEY:1968}
\begin{align}
\pdiff{P_0}{\tilde{\rho}} 
&=
\pdiff{^2 U_0}{ \tilde{Z}^2}
\\
-\frac{P_0}{\tilde{\rho}}
&=
\pdiff{^2 V_0}{ \tilde{Z}^2}
\\
\pdiff{P_0}{\tilde{Z}}
&=
0.
\end{align}
The solution of $U_0$ and $V_0$ is expressed by $P_0(\tilde{\rho})$ and the
condition of incompressibility leads to the following Reynolds equation
\begin{align}
\tilde{\rho}^2 P_0''
+ C[\lambda,\tilde{\rho}] P_0'
- P_0 + D [\lambda,\tilde{\rho}]
&=0
\label{janus.nearfield.Reynolds}
\end{align}
where $C [\lambda,\tilde{\rho}]$ and $D [\lambda,\tilde{\rho}]$ are
functions of $\tilde{\rho}$ with the parameter $\lambda$ and the boundary
conditions \NY{eq.~}(\ref{SM.3D.near.asym.BC}).
The parameter $\lambda$ is dependent on the curvature of the second
particle (See fig.~\ref{fig.sm.coordinate}).
For pairs of particles of the same size, $\lambda=-1$.

The relative motion gives singular terms arising from passive problems (a), (c),
and (g).
The passive problem (g) has only a singular torque and its force is
$\mathcal{O}(\epsilon^0)$.
As in the previous section, we evaluate the
$\mathcal{O}(\epsilon^0)$ terms using the passive problems (e) and (g).
The passive force and torque on each sphere are
\begin{align}
\frac{F_1^{(p)} }{6 \pi \eta R}
\simeq &
\frac{1}{6}
\log \epsilon \left[
\left(
u_x^{(1)}
- u_x^{(2)}
\right)
- R 
\left(
\omega_y^{(1)}
+ \omega_y^{(2)}
\right)
\right]
\nonumber \\
 &
 - 
\frac{1}{2}
\left(
u_x^{(1)}
+ u_x^{(2)}
\right)
\mathcal{I}_x^{(p)}
- 
\frac{R}{2}
\left(
\omega_y^{(1)}
- \omega_y^{(2)}
\right)
\mathcal{J}_x^{(p)}
\\
\frac{F_2 ^{(p)} }{6 \pi \eta R}
\simeq &
\frac{1}{6}
\log \epsilon \left[
-
\left(
u_x^{(1)}
-  u_x^{(2)}
\right)
+ R
\left(
\omega_y^{(1)}
+  \omega_y^{(2)}
\right)
\right]
\nonumber \\
 &
 - 
\frac{1}{2}
\left(
u_x^{(1)}
+ u_x^{(2)}
\right)
\mathcal{I}_x^{(p)}
- 
\frac{R}{2}
\left(
\omega_y^{(1)}
- \omega_y^{(2)}
\right)
\mathcal{J}_x^{(p)}
\\
\frac{T_1^{(p)} }{8 \pi \eta R^2}
 \simeq &
\frac{1}{8}
\log \epsilon \left[
- \left(
u_x^{(1)}
-  u_x^{(2)}
\right)
+ 
\frac{R}{10} \left(
4 \omega_y^{(1)}
+  \omega_y^{(2)}
 \right)
 \right.
 \nonumber \\
 &
 +
  \left.
\frac{3 R}{5}
\left(
 \omega_y^{(1)}
-  \omega_y^{(2)}
\right)
 \right]
- 
\frac{1}{2}
\left(
u_x^{(1)}
+ u_x^{(2)}
\right)
\mathcal{G}_x^{(p)}
\\
\frac{ T_2^{(p)} }{8 \pi \eta R^2}
 \simeq &
\frac{1}{8}
\log \epsilon \left[
- \left(
u_x^{(1)}
- u_x^{(2)}
\right)
+ 
\frac{R}{10} \left(
 \omega_y^{(1)}
+ 4  \omega_y^{(2)}
 \right)
 \right.
 \nonumber \\
 &
 -
 \left.
\frac{3 R}{5}
\left(
 \omega_y^{(1)}
-  \omega_y^{(2)}
\right)
\right]
 - 
\frac{1}{2}
\left(
u_x^{(1)}
+ u_x^{(2)}
\right)
\mathcal{G}_x^{(p)}
.
\end{align}
Here the constant is known for the passive problem (e) and (g) as
$
\mathcal{I}_x^{(p)} 
\simeq
0.72426
$, $
\mathcal{G}_x^{(p)} 
\simeq
0.11843
$, and $
\mathcal{J}_x^{(p)} 
\simeq
0.15802
$.

For the active problem, the force generated by the slip velocity is
\begin{align}
\frac{ F^{(a)}_1 }{6 \pi \eta R}
 \simeq & 
 \frac{1}{6} \log \epsilon
\sum_{l,m=\pm 1} 
 m
 \mathcal{N}_{l |m|}
 V_{lm}
 \nonumber \\
 &
 -
 \frac{1}{3}
\sum_{m= \pm 1}
 m \mathcal{N}_{1,|m|}
W_{1,m}
 \mathcal{I}_x^{(a)}
 \label{nonaxis.Fx1}
\\
\frac{F_2^{(a)}}{6 \pi \eta R} 
 \simeq & 
-  \frac{1}{6} \log \epsilon
\sum_{l,m=\pm 1} 
m
 \mathcal{N}_{l|m|}
 V_{lm}
 \nonumber \\
 &
 -
 \frac{1}{3}
\sum_{m= \pm 1}
 m\mathcal{N}_{1,|m|}
W_{1,m}
 \mathcal{I}_x^{(a)}
\\
\frac{ T_1^{(a)} }{8 \pi \eta R^2}
\simeq &
- \frac{1}{20} \log \epsilon
\sum_{l,m=\pm 1} 
 m \mathcal{N}_{l|m|} \left(
\frac{5}{2}V_{lm} + \frac{3}{2} W_{lm}
 \right)
  \label{nonaxis.Ty1}
\\
\frac{ T_2^{(a)} }{8 \pi \eta R^2}
\simeq &
-  \frac{1}{20} \log \epsilon
\sum_{l,m=\pm 1} 
 m \mathcal{N}_{l|m|} \left(
 \frac{5}{2}V_{lm} - \frac{3}{2} W_{lm}
\right)
.
\end{align}
The $\mathcal{O}(\epsilon^0)$ terms (the second line in the expression for the forces) are
added assuming that these forces are similar to those of an isolated
squirmer with a coefficient $\mathcal{I}_x^{(a)}$.

\section{Two Dimensions}
\label{sec.app.2D}

\subsection{Isolated squirmer}

The solution of the Stokes equation is decomposed into the radial ${\bf
r}$ and the tangential ({\bf t}) directions
\begin{align}
{\bf v}
 &=
 v_r {\bf r} + v_t {\bf t}
\end{align}
where by defining
\begin{align}
 {\bf v}_m
 &=
 \left(
v_{s,m}, -\tilde{v}_{s,m}
 \right)
 =
  v_m
 \left(
\cos m\beta, \sin m \beta
 \right)
 ,
\end{align}
\begin{align}
v_r
 =&
 {\bf u} \cdot {\bf r}
 \left(
\frac{R}{r}
 \right)^2
 +
 \sum_{m=2}^{\infty}
 \left[
 \left( \frac{R}{r} \right)^{m-1}
 - \left( \frac{R}{r} \right)^{m+1}
 \right] \frac{m}{2}
 {\bf v}_m \cdot {\bf r}_m
 \\
 v_t
 =&
-  {\bf u} \cdot {\bf t}
 \left(
\frac{R}{r}
 \right)^2
 \nonumber \\
 &
 -
 \sum_{m=2}^{\infty}
 \left[
\frac{m-2}{2}
 \left( \frac{R}{r} \right)^{m-1}
 -
\frac{m}{2}
 \left( \frac{R}{r} \right)^{m+1}
 \right] 
 {\bf v}_m \cdot {\bf t}_m
\end{align}
with
\begin{align}
 {\bf r}_m
 &=
 \left(
\cos m \theta, \sin m \theta
 \right)
 \\
 {\bf t}_m
 &=
 \left(
-\sin m \theta, \cos m \theta
 \right)
 .
\end{align}
We may define the polarity of the particle
\begin{align}
{\bf p}
 &=
 \left(
\cos \beta, \sin \beta
 \right)
\end{align}
then we may rewrite as
\begin{align}
{\bf u}
 &=
 u {\bf p}
 .
\end{align}

\subsection{Far-field interaction}

\tlll{As in the three dimensional case, we use Faxen's laws to evaluate the 
far-field interaction.}
Since we are working in a force-free system, the leading order term
should be ${\bf u} = {\bf v}_0 + \mathcal{O} (\nabla^2 {\bf v}_0)$.
The motion of one particle perturbed by a  second particle is
\begin{align}
{\bf u}^{(i)}
 = &
 u {\bf p}^{(i)}
 +
 \sum_{j \neq i}
 \left( \frac{R}{r_{ij}} \right)^2
 \left[
\left(
 {\bf u}^{(j)} \cdot \hat{\bf r}_{ij}
 \right) \hat{\bf r}_{ij}
 -
 \left(
 {\bf u}^{(j)} \cdot \hat{\bf t}_{ij}
 \right) \hat{\bf t}_{ij}
 \right]
\nonumber \\
& +
 \sum_{j}
 \frac{R}{2r_{ij}}
 \left( {\bf v}^{(j)}_2 \cdot \hat{\bf r}_{ij,2} \right) \hat{\bf r}_{ij}
\end{align}
where
\begin{align}
\hat{\bf r}_{ij}
 &=
 \frac{{\bf r}^{(i)} - {\bf r}^{(j)}}{r_{ij}}
 =
 \left(
\cos \theta_{ij}, \sin \theta_{ij}
 \right)
 \\
\hat{\bf r}_{ij,m}
 &=
 \left(
\cos m \theta_{ij}, \sin m \theta_{ij}
 \right)
\end{align}
and the unit tangent vector is
\begin{align}
\hat{\bf t}_{ij}
 &=
 \left(
 -\sin \theta_{ij}, \cos \theta_{ij}
 \right)
 .
\end{align}
We may define the generalized tangent vector as 
\begin{align}
\hat{\bf t}_{ij,m}
 &=
 \left(
 -\sin m\theta_{ij}, \cos m\theta_{ij}
 \right)
 ,
\end{align}
these vectors are transformed by $\theta_{ij} \leftrightarrow
\theta_{ji}=\theta_{ij}+\pi$ as ${\bf r}_{ij,m} = (-1)^m {\bf r}_{ji,m}$
and ${\bf t}_{ij,m} = (-1)^m {\bf t}_{ji,m}$.
The angular velocity is
\begin{align}
{\bf \omega}^{(i)}
 &=
 -\frac{1}{4}
 \sum_{j \neq i}
 ({\bf v}_m^{(j)} \cdot \hat{\bf t}_{ij} )
 \frac{R}{r_{ij}^2} {\bf e}_z
\end{align}
Note that even in two dimensions, there is no contribution from the
$m=1$ mode on the rotation, because the flow field of the $m=1$ mode is in fact the same as
potential flow.
\NYY{
The explicit forms in \NY{eqs.~}(\ref{trans.vel.2D}) and (\ref{ang.vel.2D})
\begin{align}
 u_{\parallel,1}
 &=
 \left( \frac{R}{r_{ij}} \right)^2
\left(
 {\bf u}^{(j)} \cdot \hat{\bf r}_{ij}
 \right) 
\\
 u_{\perp,1}
 &=
 -
  \left( \frac{R}{r_{ij}} \right)^2
 \left(
 {\bf u}^{(j)} \cdot \hat{\bf t}_{ij}
 \right)
 \\
  u_{\parallel,2}
  &=
 \frac{R}{2r_{ij}}
 \left( {\bf v}^{(j)}_2 \cdot \hat{\bf r}_{ij,2} \right) \hat{\bf r}_{ij}
\\
 \omega_2
 &=
 -\frac{1}{4}
 ({\bf v}_2^{(j)} \cdot \hat{\bf t}_{ij} )
 \frac{R}{r_{ij}^2}
\end{align}
}

\subsection{Near-field interaction}

The lubrication theory of  a cylinder approaching toward a wall or another
cylinder was discussed in \cite{Jeffery:1922a,Jeffrey:1981}.
The analysis was first done using bipolar coordinates \cite{Jeffery:1922a}
It has been further extended to the flow around a cylinder in a
confined space \cite{Yang:2013,Cardinaels:2015}.
In this section, we will use the same stretched coordinates as
three-dimensional flow.

We may divide the motion into components of the parallel and perpendicular directions to the
center line.
In stretched coordinates, the slip velocity is
\begin{align}
 {\bf v}_s^{(i)}
 &=
 v_s^{(i)} {\bf t}^{(i)},
\end{align}
which is expanded in $\epsilon$ in a similar fashion to the three dimensional case.
The superscript $i=1,2$ denotes the sphere 1 or 2.
In stretched coordinates, we expand the velocity field and pressure.
As in the three-dimensional case, the velocity and angular velocity
are calculated from the force-free and torque-free conditions.
From the  linearity of the problem, we may decompose $u_y$ and $(u_x, \omega)$. 
By rotating the result, the velocity and angular velocity in general
coordinates are given by
\begin{align}
 {\bf u}^{(i)}
 &=
 u_{\perp} \hat{\bf t}_{ji}
 +
 u_{\parallel} \hat{\bf r}_{ji}
 \\
  \omega^{(i)}
  =&
 -
 \frac{1}{4}
 \left(
{\bf u}_0^{(i)} - {\bf u}_0^{(j)}
 \right)
 \cdot
 \hat{\bf t}_{ji}
 \nonumber \\
 &
 +
 \frac{1}{4}
 \sum_m
  \left(
3 (-1)^{m-1}
 {\bf v}^{(i)}_{s,m}
 + {\bf v}^{(j)}_{s,m}
 \right) \cdot \hat{\bf t}_{ji,m}
 .
\end{align}
Here we define the unit vectors with ${\bf r}_{21} = {\bf r}_2 - {\bf r}_1$.
The concrete form is
\begin{align}
 u_{\parallel}
  = &
 \frac{1}{2}
 \left(
{\bf u}_0^{(i)} + {\bf u}_0^{(j)}
 \right)
 \cdot
 \hat{\bf r}_{ji}
 \nonumber \\
 &
 -
 \frac{1}{2} \sqrt{\epsilon}
 \sum_m
 m \left(
 (-1)^m
 {\bf v}^{(i)}_{s,m}
 + {\bf v}^{(j)}_{s,m}
 \right) \cdot \hat{\bf r}_{ji,m}
\\
 u_{\perp}
  = &
 \frac{1}{4}
 \left(
3 {\bf u}_0^{(i)} + {\bf u}_0^{(j)}
 \right)
 \cdot
 \hat{\bf t}_{ji}
 \nonumber \\
 &
 -
 \frac{1}{4}
 \sum_m
  \left(
 (-1)^m
 {\bf v}^{(i)}_{s,m}
 + {\bf v}^{(j)}_{s,m}
 \right) \cdot \hat{\bf t}_{ji,m}
 .
\end{align}
\NYY{
The explicit forms in \NY{eqs.~}(\ref{trans.vel.2D}) and (\ref{ang.vel.2D}) are as follows:
\begin{align}
 u_{\parallel,1}
  = &
 \frac{1}{2}
 \left(
{\bf u}_0^{(i)} + {\bf u}_0^{(j)}
 \right)
 \cdot
 \hat{\bf r}_{ji}
 -
 \frac{1}{2} \sqrt{\epsilon}
  \left(
 -
 {\bf v}^{(i)}_{s,1}
 + {\bf v}^{(j)}_{s,1}
 \right) \cdot \hat{\bf r}_{ji,1}
 \\
 u_{\parallel,m}
 &=
  -
 \frac{1}{2} \sqrt{\epsilon}
 m \left(
 (-1)^m
 {\bf v}^{(i)}_{s,m}
 + {\bf v}^{(j)}_{s,m}
 \right) \cdot \hat{\bf r}_{ji,m}
\\
 u_{\perp,1}
  = &
 \frac{1}{4}
 \left(
3 {\bf u}_0^{(i)} + {\bf u}_0^{(j)}
 \right)
 \cdot
 \hat{\bf t}_{ji}
 -
 \frac{1}{4}
  \left(
 -
 {\bf v}^{(i)}_{s,1}
 + {\bf v}^{(j)}_{s,1}
 \right) \cdot \hat{\bf t}_{ji,1}
 \\
 u_{\perp,m}
 &=
 -
 \frac{1}{4}
  \left(
 (-1)^m
 {\bf v}^{(i)}_{s,m}
 + {\bf v}^{(j)}_{s,m}
 \right) \cdot \hat{\bf t}_{ji,m}
 \\
   \omega_1
  =&
 -
 \frac{1}{4}
 \left(
{\bf u}_0^{(i)} - {\bf u}_0^{(j)}
 \right)
 \cdot
 \hat{\bf t}_{ji}
  +
 \frac{1}{4}
  \left(
3 
 {\bf v}^{(i)}_{s,1}
 + {\bf v}^{(j)}_{s,1}
 \right) \cdot \hat{\bf t}_{ji,1}
 \\ 
  \omega_m
  =&
 \frac{1}{4}
  \left(
3 (-1)^{m-1}
 {\bf v}^{(i)}_{s,m}
 + {\bf v}^{(j)}_{s,m}
 \right) \cdot \hat{\bf t}_{ji,m}
 .
\end{align}
}


\end{document}